\documentclass[prb, aps, twocolumn, 10pt, showpacs, final, floatfix]{revtex4-1}
  
\pdfoutput=1

\usepackage{graphicx}
\usepackage{amsmath, amssymb, amsfonts, bm}
\usepackage{bbm}
\usepackage{latexsym}
\usepackage[colorlinks]{hyperref}
\usepackage{mathdots}
\usepackage[protrusion=true,expansion=true]{microtype}

\newcommand{\matr}[1]{\mathsf{#1}}

\newcommand{\ee}{\mathrm{e}}
\newcommand{\ii}{\mathrm{i}}
\newcommand{\D}[1]{\,\text{d}#1\,}

\newcommand{\bra}[1]{\mathinner{\langle{#1}|}}
\newcommand{\ket}[1]{\mathinner{|{#1}\rangle}}
\newcommand{\inner}[3]{\left< #1\left|#2\right|#3\right >}
\newcommand{\braket}[2]{\langle #1 | #2 \rangle}

\newcommand{\expe}[1]{\left\langle #1 \right\rangle}

\begin{document}
%--------------------------------%
\title{Probe spectroscopy of quasienergy states}
\author{Matti Silveri}
\author{Jani Tuorila}
\author{Mika Kemppainen}
\author{Erkki Thuneberg}
\affiliation{Department of Physics, University of Oulu, P.O. Box 3000, FI-90014, Finland}

\date{\today}
%--------------------------------%

%--------------------------------%
\begin{abstract}
The present qubit technology, in particular in Josephson qubits, allows an unprecedented control of discrete energy levels.  This motivates a new study of the old pump-probe problem, where a discrete quantum system is driven by a strong drive and simultaneously probed by a weaker one.  The strong drive is included by the Floquet method and the resulting quasienergy states are then studied with the probe. We study a qubit where the harmonic drive has a significant longitudinal component relative to the static equilibrium state of the qubit. Both analytical and numerical methods are used to solve the problem. We present calculations with realistic parameters and compare the results with recent experimental results. A short introduction to the Floquet method and the probe absorption is given.  
\end{abstract}
%--------------------------------%
\pacs{03.65.Sq, 31.15.xg, 85.25.Cp, 42.50.Ct}
%03.65.Sq Quantum mechanics; semiclassical theories and application
%31.15.xg Calculations and mathematical methods in atomic and molecular physics; semiclassical methods
%85.25.Cp Superconducting devices; Josephson devices
%42.50.Ct Quantum optics; Quantum description of interaction of light and matter; related experiments
\maketitle

%--------------------------------%
%--------------------------------%
\section{Introduction}
The discrete energy levels of a quantum system can be mapped by studying the absorption from a weak harmonic perturbation. A coupling with a monochromatic drive changes the characteristics of the spectrum, an effect known  as the dynamic Stark shift~\cite{Autler55}. In atomic physics, this has a prototype in the form of an atom, driven with one laser, the pump, and probed with another of low intensity~\cite{Mollow72, Wei97} (see Fig.~\ref{TwoTone.fig}). Instead of the bare atomic levels, the probe induces transitions between the dressed states of the coalesced atom and pump. 

The standard case in atomic physics is a dipole transition, where the strong drive is transverse to the static Hamiltonian. Similarly in nuclear magnetic resonance, the drive field is transverse to the static field. This limitation has been removed by the introduction of new systems where longitudinal drive can be included. Examples are Rydberg states of an atom~\cite{Noel98}, Josephson qubits~\cite{Oliver05, Sillanpaa06, Wilson07, Tuorila10, Izmalkov08}, nitrogen vacancy centers in diamond~\cite{Childress10}, and semiconductor quantum-dots~\cite{Petta10,Gaudreau12, Petta12}. This motivates a renewed study  of the pump-probe physics.

The purpose of this paper is to present calculations on pumped and probed qubits, where the pump has an essential longitudinal component with respect  to the equilibrium level splitting. We start by a brief presentation of the Floquet formalism that is used to calculate the dynamic Stark effect of the pump field (Sec.~\ref{Floquet.subsec}). The formalism is enjoying a revival due to the generation of novel systems allowing strong driving fields and long enough coherence times, e.g. superconducting qubits and circuits~\cite{Son09, Bushev10, Ferron10, Leppakangas10, Tuorila10, Hausinger10, Russomanno11, Satanin12, Sauer12}, and topological insulators~\cite{Lindner11}. The Floquet formalism~\cite{Shirley65, GrifoniHnggi98,Chu04} is a semiclassical method that can be understood as a limiting case of a full quantum mechanical picture when the number of quanta in the driving field is large~\cite{Shirley65}. The resulting dressed states are called quasienergy states. We sketch the theory of weak probe absorption and dispersion on the quasienergy states (Sec.\ \ref{Probe_sec}). This method has been used in the literature \cite{Madison98,Breuer88, Sauer12} but, to our knowledge, has not been properly justified. There has been many experiments in the field of superconducting Josephson qubits~\cite{Sillanpaa06, Wilson07, Gunnarsson08, Izmalkov08} which could have been interpreted in terms of the probe absorption spectroscopy of the quasienergy levels. Nevertheless, only one measurement has invoked the method~\cite{Tuorila10}. The general theory is applied to two level systems (Sec.\ \ref{Appl_sec}). We compare numerical calculations with analytical  approximations.  We present calculations with parameter values that are relevant for recent experiments~\cite{Wilson07, Izmalkov08}, and compare the calculated spectra with the measured ones.

The Floquet approach used here should be compared with an alternative method of solving the pump-probe problem. The conventional approach\cite{Mollow72, Wei97} is to first find the steady state solution of the density matrix corresponding to the driven, but not probed, system. Then the probe absorption rate is obtained by calculating a correlator of the probe Hamiltonian at the probe frequency~\cite{Mollow72}. The two methods are identical with the exception that the approximations concerning relaxation  can be different. The Floquet approach has the benefit that the quasienergy structure gives additional insight and allows a simple analysis of the probe absorption by Fermi's golden rule.  

%--------------------------------%
%--------------------------------%
\section{Quasienergy states}\label{Floquet.subsec}
\begin{figure}
\includegraphics[width=0.7\columnwidth]{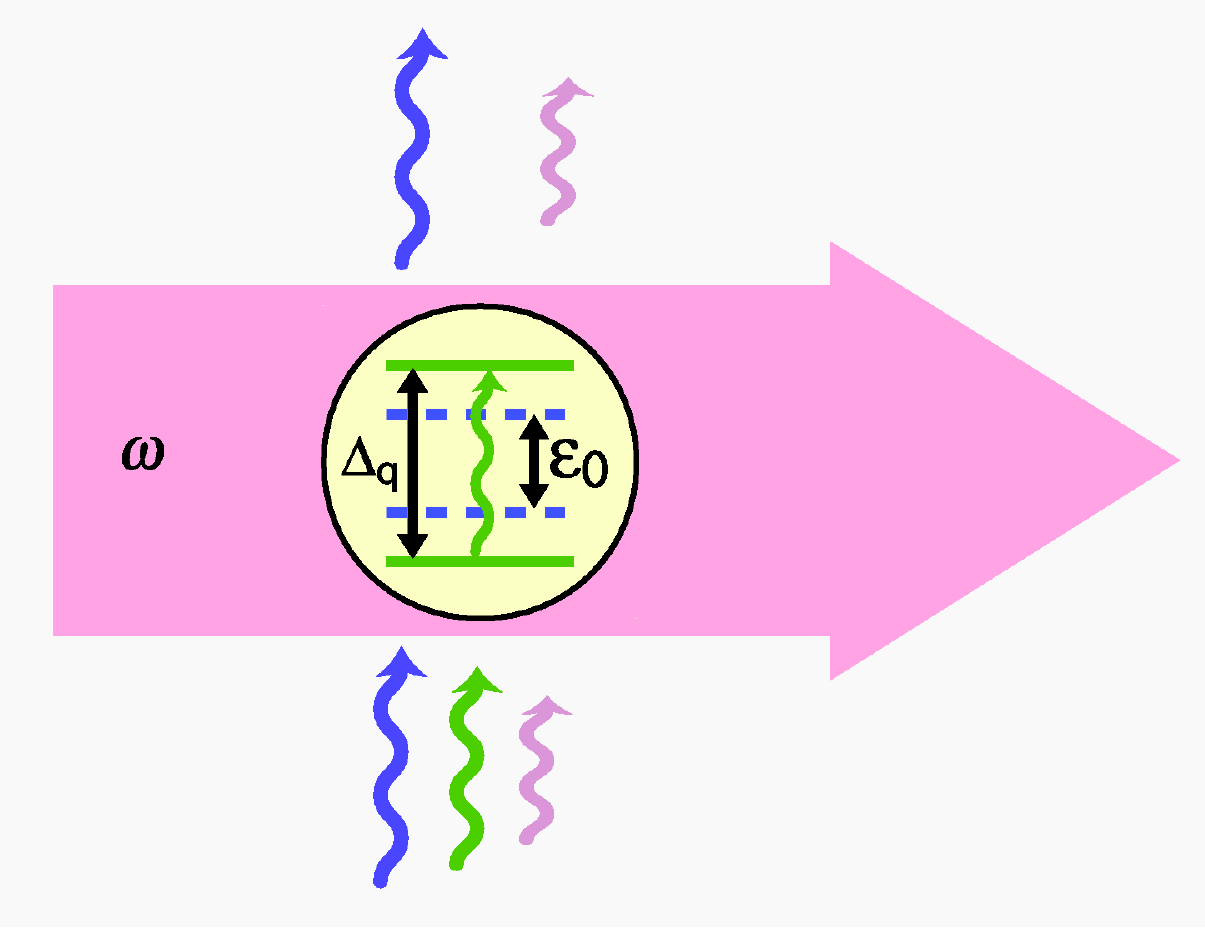}
\caption{\label{TwoTone.fig}Illustration of the absorption spectrum of a strongly driven atom. The atom with an energy separation $\varepsilon_0$ is driven with an angular frequency $\omega=2\pi/\tau$. The drive changes the atomic energy (quasienergy $\Delta_{\rm q}$) leading to a shift of the resonance peak in the absorption spectrum (dynamic Stark shift).}
\end{figure}

We study a driven system described by the Hamiltonian $\hat{H}(t)$
\begin{equation}
  \hat{H}(t)=\hat{H}_0+\hat{H}_{\rm{S}}(t) \label{totalHam.eq}.
\end{equation}
The time-independent $\hat{H}_0$ represents the atomic system expressed in an atomic basis $\mathcal{B}_{\rm A} =\{|\sigma\rangle\} $, spanning the atomic Hilbert space $\mathcal{H}_{\rm A}$. The time-dependent term is $\tau$-periodic, $\hat{H}_{\rm{S}}\left(t+\tau\right)=\hat{H}_{\rm{S}}(t)$, and represents the strong driving of the atomic system. The effect of the strong driving can be seen as a change in the energy level structure of the atom. By studying the absorption profile of the driven atom (Sec.\ \ref{Probe_sec}), one observes that the locations of the spectral lines move as a function of the drive intensity. This  dynamic Stark effect~\cite{Autler55} is depicted in Fig.~\ref{TwoTone.fig}.

%--------------------------------%
%--------------------------------%
 
The time-dependent Schr\"odinger equation is\begin{equation}
\left(-\ii\hbar\frac{\D{}}{\D{t}}+\hat{H}(t)\right)\ket{\Psi(t)}=0. \label{SE.eq}
\end{equation}
Due to the periodicity of $\hat{H}(t)$, this is now analogous to the one-dimensional Bloch's problem of solid state physics~\cite{AM}. Within the analogy, the solution of Eq.~\eqref{SE.eq} can be expressed in the form
\begin{equation}
\ket{\Psi(t)}=\ee^{-\ii\epsilon t/\hbar} \ket{u(t)}, \label{BlochAnsatz.eq}
\end{equation}
where the state $\ket{u(t)}$ is $\tau$-periodic and $\epsilon$ is called the quasienergy~\cite{Shirley65}.

In order to solve Eq.~\eqref{SE.eq} by using~\eqref{BlochAnsatz.eq}, the Floquet method takes advantage of the periodicity of $\hat{H}(t)$ and $\ket{u(t)}$. The atomic Hilbert space $\mathcal{H}_{\rm A}$ is expanded with $\tau$-periodic functions, $\mathcal{H}_{\tau}$, spanned by the basis $\mathcal{B}_{\tau}=\{\ket{n}, n\in\mathbbm{Z}; \braket{t}{n}=\exp(\ii n \omega t) \}$, where $\omega=2\pi/\tau$. This composite Hilbert space $\mathcal{H}_{\rm A}\otimes\mathcal{H}_{\tau}$ is referred to as the Sambe space~\cite{Sambe72}. This expansion allows the representation of the periodic quantities in terms of time-independent coefficients 
\begin{align}
  H_{\sigma\sigma'}^{(n,m)}&\equiv \inner{\sigma,n}{\hat{H}(t)}{\sigma',m}\notag\\
&=\frac{1}{\tau}\int_0^{\tau} \D{t}\inner{\sigma}{\hat{H}(t)}{\sigma'}\ee^{-\ii (n-m) \omega t},\label{fourierH.eq}\\
c^{(n)}_{\sigma}&\equiv \braket{\sigma,n}{u(t)}= \frac{1}{\tau}\int_0^{\tau}\D{t}\braket{\sigma}{u(t)}\ee^{-\ii n \omega t}\label{fourieru.eq}.
\end{align} 
Eqs.~\eqref{fourierH.eq} and~\eqref{fourieru.eq} can also be seen as the Fourier series representation of $H_{\sigma\sigma'}(t)$ and $\ket{u_\sigma(t)}$.

In the Sambe space, the Schr\"odinger equation~\eqref{SE.eq} reduces to 
\begin{equation}
\sum_{\sigma'}\sum_{m=-\infty}^\infty  \underbrace{\left(m\hbar\omega \delta_{\sigma\sigma'} \delta_{nm} +H^{(m-n)}_{\sigma\sigma'}\right)}_{(\matr{H}_{\rm F})^{(n,m)}_{\sigma\sigma'}} c^{(m)}_{\sigma'}=\epsilon c^{(n)}_{\sigma}, \label{Floquet.eq}
\end{equation}
which can be understood as a time-independent (Floquet matrix) eigenvalue problem
\begin{equation}
\matr{H}_{\rm F}\ket{u}=\epsilon\ket{u} \label{FloquetMatrix.eq},
\end{equation}
with an infinite rank. The quasienergy $\epsilon$ can now be understood as the combined energy of the atom and the driving field, and it is analogous to the quasimomentum in solid state physics~\cite{AM}. Quasienergy states $\ket{u}=\sum_{\sigma n}c^{(n)}_{\sigma}\ket{\sigma,n}$ are obtained as the eigenvectors of the Hermitian matrix $\matr{H}_{\rm F}$, and thus form a complete and orthonormal basis of the Sambe space. Generally, the eigenvalue equation~\eqref{FloquetMatrix.eq} has to be solved numerically by using an appropriate truncation. In some limits approximate analytic solutions can be found (see Sec.~\ref{subsec.choiceofbasis}-\ref{subsec.Quasienergies}). It is worthwhile to notice the beauty of the Floquet approach, it reduces time-dependent problems to static ones, that is, to time-independent eigenvalue problems.

The quasienergies obtained from the eiqenvalue equation (\ref{FloquetMatrix.eq}) have a periodic structure. Corresponding  to a quasienergy $\epsilon_r$, there is an infinite set of solutions with quasienergies $\epsilon_{r,n}=\epsilon_{r}+n\hbar\omega$, where $n$ is an integer. The corresponding eigenstates $\ket{u_{r,n}}$ are trivially obtained from each other as they produce the same $\ket{\Psi(t)}$ in Eq.\ (\ref{BlochAnsatz.eq}). Therefore it is sufficient to solve numerically the states in a single quasienergy interval of width $\hbar\omega$, which is referred to as a single Brillouin zone. The number of such states equals the number of basis states of the atomic Hamiltonian $\hat{H}_0$. Physically the quasienergy states can be interpreted as the atomic states being entangled with the driving field containing different number of quanta. 

%--------------------------------%
\section{Probe spectroscopy of quasienergy levels}\label{Probe_sec}

The quasienergy spectrum can be studied in terms of absorption from a weak perturbation, similar to the time-independent quantum systems. As a demonstration of the power of the Floquet method, we derive the transition rate between two quasienergy states in a similar fashion to Fermi's golden rule. However, we generalize the time-independent results, typically derived by using harmonic perturbation, by allowing the probe Hamiltonian to be quasiperiodic
\begin{equation}
\hat{H}_{\rm P}(t)=F_{\rm P}(t)\hat{F}_{\rm S}(t)+\left[F_{\rm P}(t)\hat{F}_{\rm S}(t)\right]^\dagger.\label{probeProduct.eq}
\end{equation} 
Here, $F_{\rm P}$ is $\tau_{\rm P}$-periodic and $\hat{F}_{\rm S}$ is $\tau$-periodic. When the periods are incommensurate, the probe Hamiltonian $\hat{H}_{\rm P}(t)$ is not periodic, in spite of consisting of products of periodic operators. The quasiperiodic form (\ref{probeProduct.eq}) allows various realizations of the probe~\cite{Tuorila10, Wilson07, Izmalkov08}. We take the $\tau_{\rm P}$-periodic part of the probe~\eqref{probeProduct.eq} to have the harmonic form 
$F_{\rm P}(t)=A_{\rm P}\ee^{-\ii\omega_{\rm P} t} $, with the amplitude $A_{\rm P}$ and the angular frequency $\omega_{\rm P}=2\pi/\tau_{\rm P}$. More general functions can be decomposed into Fourier series, where each term can be treated independently of the others. The $\tau$-periodic part of the probe becomes time-independent in the Sambe space, and can be represented in the quasienergy basis as
\begin{equation}
\hat{F}_{\rm S}=\sum_{p,q} F_{pq} \ket{u_p}\bra{u_q}, \label{perturbationTermExp.eq}
\end{equation}
where the summations go over all quasienergy states (all Brillouin zones). The matrix elements
\begin{equation}
F_{pq}\equiv \inner{u_{p}(t)}{\hat{F}_{\rm S}(t)}{u_{q}(t)} = \inner{u_{p}}{\hat{F}_{\rm S}}{u_{q}}.   \label{Numerical.pfi}
\end{equation}
do not depend on time and are easy to implement after the numerical solution of the Floquet eigenvalue problem~\eqref{FloquetMatrix.eq}. 

\subsection{Golden rule for transitions between quasienergy state}
We include the probe Hamiltonian~\eqref{probeProduct.eq} into the Schr\"odinger equation~\eqref{SE.eq}, and look for a solution in the form
\begin{equation}
  \ket{\Psi(t)}=\sum_{q} a_q(t) \ee^{-\ii\epsilon_{q}t/\hbar}\ket{u_{q}},\label{Ansatz.eq}
\end{equation}
where the summation goes over all quasienergy states. This leads into a differential equation for the probability amplitude $a_{f}(t)$ to be in the state $f$
\begin{align}
\dot{a}_{f}(t)&=-\frac{\ii A_{\rm P}}{\hbar}\sum_{i}a_i(t)\ee^{\ii(\omega_{fi}-\omega_{\rm P})t}\inner{u_{f}}{\hat{F}_{\rm S}}{u_{i}}\notag\\ &-\frac{\ii A_{\rm P}^*}{\hbar}\sum_{i}a_i(t)\ee^{\ii(\omega_{fi}+\omega_{\rm P})t}\inner{u_{f}}{\hat{F}^\dagger_{\rm S}}{u_{i}}, \label{timeRateA.eq}
\end{align}
where $\omega_{fi}=(\epsilon_{f}-\epsilon_{i})/\hbar$ denotes the transition frequency between the quasienergies.

We now assume that the probe amplitude $A_{\rm P}$ is small. When the matrix elements $F_{fi}$ are small in comparison to the characteristic energy quantum $\hbar\omega_{\rm P}$, it is sufficient to make a perturbative expansion in Eq.~\eqref{timeRateA.eq} up to first order in $A_{\rm P}$~\cite{Sakurai}. The integration over the broadened final state leads to a finite steady  state rate $\mathcal{P}_{i\rightarrow f}$ for the transition from the initial $\ket{\Psi_i(t)}$ to the final quasienergy state $\ket{\Psi_f(t)}$. We assume that the transition rate is small in comparison to the broadening of the final state. The summation over all final quasienergy states gives the absorptive ($\omega_{fi}>0$) transition rate
\begin{equation}
  \mathcal{P}=\frac{|A_{\rm P}|^2}{\hbar^2}\sum_{ i,f}p_i \frac{\gamma_{fi} \left|\inner{u_{f}}{\hat{F}_{\rm S}}{u_{i}}\right|^2}{(\omega_{fi}-\omega_{\rm P})^2+\frac14\gamma_{fi}^2}. \label{goldenrule.sum.eq}
\end{equation}
where $\gamma_{fi}=\gamma_i+\gamma_f$ is the sum of the widths of the initial and finals states, which both are assumed to have Lorentzian form.  We have also summed over all  initial quasienergy states weighted by their  steady state occupation probabilities $p_i$.
We have neglected all but the resonant term in Eq.~\eqref{timeRateA.eq}. This is analogous to the rotating wave approximation (RWA) where the rapidly oscillating terms are assumed to average out in the steady state dynamics.  

The result~\eqref{goldenrule.sum.eq} can be named as \emph{Fermi's golden rule for transitions between quasienergy states}, as it is analogous to the result obtained between energy eigenstates~\cite{Sakurai}. The transitions between the quasienergy states occur when the corresponding quasienergy difference equals the energy quantum of the probe: $\epsilon_{f}-\epsilon_i=\hbar\omega_{\rm P}$.  The magnitude of the transition is proportional to the squared matrix element $|F_{fi}|^2$. It is worthwhile to note that the transition does not occur between the atomic states $\ket{\sigma}$, but between the quasienergy states $\ket{u}$. As seen by the atomic system, multiple strong driving quanta can participate in the process since the quasienergy states can be in any Brillouin zone: $\epsilon_{f0}-\epsilon_{i0}+(m-n)\hbar\omega=\hbar\omega_{\rm P}$. Yet, within the first order approximation and with harmonic perturbation,  only one probe quantum can be exchanged in the transition process. 

\subsection{Relation to the spectrum of the probe field} \label{sec:inoutput}

The transition rate $\mathcal{P}$ can be expressed alternatively using correlation functions. By applying the completeness of the quasienergy states $\ket{u_{r}}$ one finds that $\mathcal{P}$~\eqref{goldenrule.sum.eq} is equal to the absorption spectrum~\cite{LandauStat} 
\begin{equation}
S(\Omega)= \frac{|A_{\rm P}|^2}{\hbar^2} \int_{-\infty}^{\infty} \expe{\hat{F}_{\rm S}^\dagger(t) \hat{F}_{\rm S}(0)} \ee^{\ii \Omega t} \D{t},\label{inoutput.spec.eq}
\end{equation}
at the probe frequency. The same expression is also obtained by formulating the probe absorption spectroscopy using the input-output-formalism~\cite{QuantumNoise,GardinerCollett}. In the corresponding emission spectrum, the order of the operators in the correlator is interchanged. The correlator approach~\eqref{inoutput.spec.eq}, usually calculated using numerical integration of the master equation, gives the same information about the locations and widths of the spectrum peaks as the Floquet approach~\eqref{goldenrule.sum.eq}. Nevertheless, the possible resonance shifts (i.e., Stark~\cite{Autler55} and Bloch-Siegert~\cite{BS} shifts) or the magnitudes of the resonances are cleanly explained in the Floquet method with the quasienergy structure~\cite{Tuorila10} giving physical insight on the composition of the driving field and the system.

In the linear response theory, the absorption spectrum is given by the imaginary part of the generalized probe susceptibility $\alpha(\omega_{\rm P})=\alpha'(\omega_{\rm P})+\ii \alpha''(\omega_{\rm P})$~\cite{LandauStat}, a function which determines the dynamics of the system under perturbation. This is generally referred to as the fluctuation-dissipation theorem. Thus, the imaginary part $\alpha''(\omega_{\rm P})$ becomes directly proportional to the absorption $\mathcal{P}(\omega_{\rm P})$~\eqref{goldenrule.sum.eq} of the system. Moreover, the real part $\alpha'(\omega_{\rm P})$ (dispersion), which makes the phase shift of the response, can be obtained analytically from $\alpha''(\omega_{\rm P})$ [in practice Eq.~\eqref{goldenrule.sum.eq}] by using the Kramers-Kronig -relations. This way one can solve both the absorption and the dispersion by exploiting the quasienergy structure.

\subsection{Extensions}
The golden rule~\eqref{goldenrule.sum.eq} is a perturbative result in the perturbation parameter $\lambda=F_{fi} A_{\rm P}/\hbar\omega_{\rm P}$. The second order contributions become significant when $\lambda$ is comparable to unity. With such large transition strengths, the original quasienergy structure becomes altered by the probe field. Instead of calculating the higher order expansions in $\lambda$, we propose the use of the generalized Floquet method~\cite{Chu83, Chu04}. It allows the calculation of the quasienergies of a Hamiltonian having two, or more, driving fields with arbitrary driving amplitudes $A_j$ and (incommensurate) frequencies $\omega_j$. The resulting quasienergy structure is 'quasiperiodic', which in the case of two driving fields means that $\epsilon_{r,n,m}=\epsilon_{r}+n\hbar\omega_1+m\hbar\omega_2$. 

The detailed analysis of the two-mode quasienergy structure provides a quantitative method to study, among others, the validity limit of the first order expansion leading to the golden rule~\eqref{goldenrule.sum.eq}. A comparison can be made by studying the differences between the quasienergy levels calculated with and without the probe field. One can say that the golden rule consideration is not valid if the results differ outside the expected locations of the weak probe resonances (anti-crossings), or if these locations are shifted. Details of the generalized Floquet method applied to the strongly driven and weakly probed quantum two-level system are given in Appendix\ref{App.two-mode}.

\section{Application to a two-level system}\label{Appl_sec}
\begin{figure}%[tbp!]
\centering
\includegraphics[height=0.75\columnwidth]{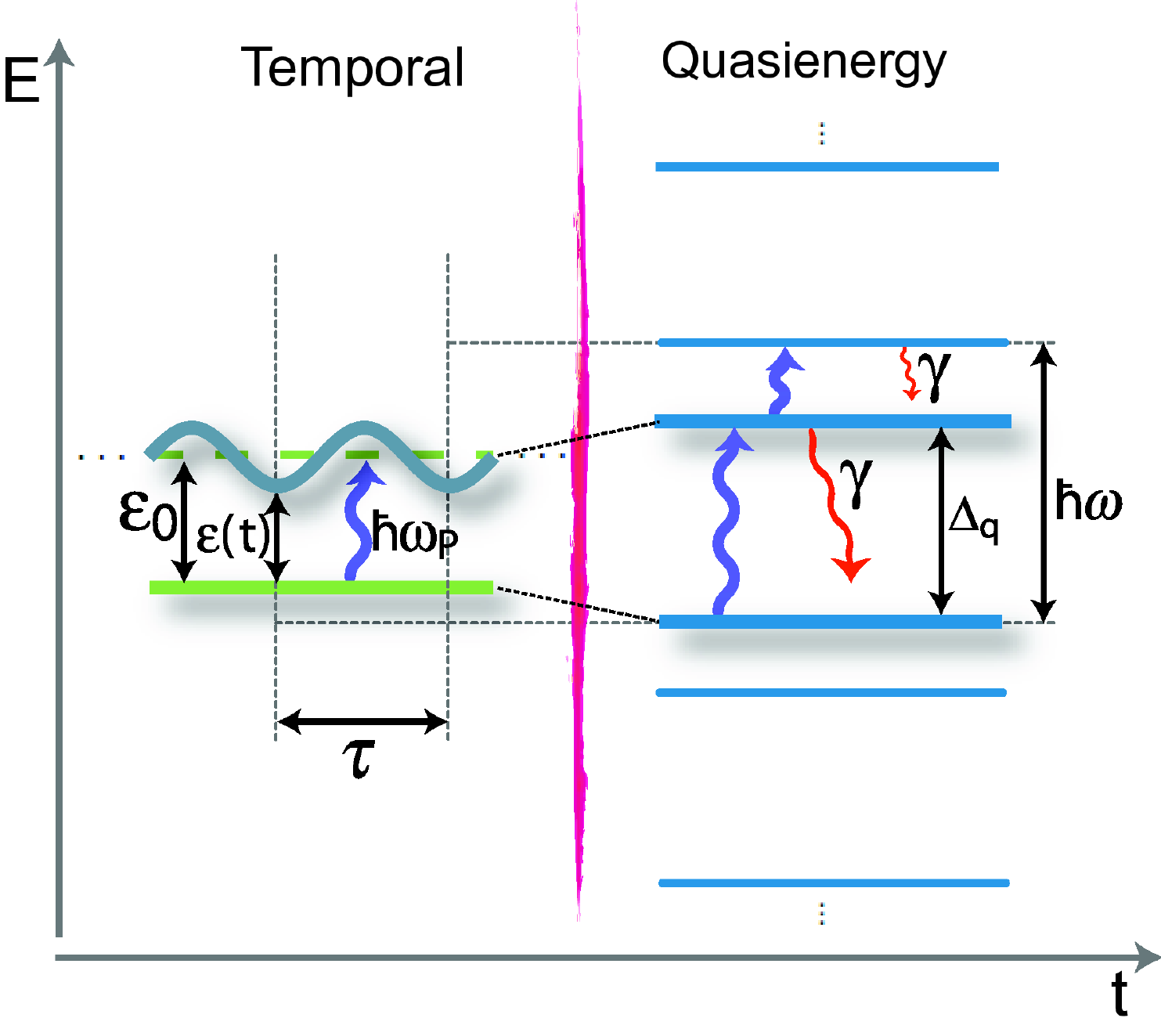}
\caption{\label{TemporalQuasienergy.fig}Schematics of the longitudinally driven and weakly probed two-level system both in the temporal space and the quasienergy space. \emph{The temporal space}; A two-level system, whose energy splitting $\varepsilon(t)$ oscillates sinusoidally around the mean $\varepsilon_0$ with the period $\tau$. \emph{The quasienergy space}; Transformation to the Floquet formalism results in the temporally static quasienergy levels that repeat in energy with the period $\hbar\omega$. The weak probe field (blue arrows) acts between the atomic states or between the quasienergy states.  In the case of $\omega_{\rm P}<\omega$, the probe resonance is met when $\Delta_{\rm q}=\hbar\omega_{\rm P}$ or $\hbar\omega-\Delta_{\rm q}=\hbar\omega_{\rm P}$.  }
\end{figure}

We give an example on the probe spectroscopy of quasienergy states by studying a two-level system under a strong longitudinal drive~\cite{Sillanpaa06, Wilson07, Gunnarsson08, Izmalkov08} (see Fig.~\ref{TemporalQuasienergy.fig}). In similar systems~\cite{Shevchenko09,Ashhab07, Son09, Hausinger10}, one has previously considered the rotating wave approximation (RWA) and the Landau-Zener-St\"uckelberg (LZS) approach~\cite{LZSM}, whose point of view is in the discretized, 'stroboscopic', evolution of the periodically driven qubit in the temporal space. In the LZS-approach, the inclusion of an additional probe field is complicated. In contrast, we concentrate on the possibility to directly map the quasienergies by studying absorption from the probe. 
%------------------------------------------%
%------------------------------------------%

We assume the Hamiltonian 
\begin{equation}
  \hat{H}(t)=\frac{\varepsilon_0}{2} \hat{\sigma}_z+\frac{\Delta}{2}\hat{\sigma}_x  +\frac{A}{2} \cos(\omega t) \hat{\sigma}_z
  +\frac{A_{\rm P}}{2} \cos(\omega_{\rm P} t)\hat{\sigma}_z , \label{general.ham}
\end{equation}
where the operators $\hat{\sigma}_{x,y,z}$ denote the Pauli spin matrices. 
Here the first two terms form the atomic part $\hat H_0$, which consists of a static energy splitting $\varepsilon_0$ and a tunneling amplitude $\Delta$. The third term is the strong drive $ \hat{H}_{\rm{S}}(t)$. Together with the first term, this implies that the level spacing $\varepsilon(t)=\varepsilon_0+A \cos(\omega t)$ (neglecting $\Delta$) oscillates  with amplitude $A$ and frequency $\omega=2\pi/\tau$. 
After transforming to the Floquet formalism, this is reflected in the periodicity in the quasienergy, see Fig.~\ref{TemporalQuasienergy.fig}. In a two-level system, we define the quasienergy splitting $\Delta_{\rm q}=\epsilon_+-\epsilon_-$ as the energy difference between the two quasienergy levels within a Brillouin zone. This, together with the $\hbar\omega$-periodicity, includes all relevant information about the energy level structure of the driven two-level system.  The fourth term in (\ref{general.ham}) is the probe Hamiltonian  $\hat{H}_{\rm P}(t)$. It is assumed to act in the same direction as the drive, but with a small amplitude~$A_{\rm P}$ and a different frequency~$\omega_{\rm P}$.
The same  direction can be arranged, e.g., by coupling the probe  to the system through the same channel as the strong drive. For simplicity, we consider here a purely $\tau_{\rm P}$-periodic probe Hamiltonian~\eqref{probeProduct.eq} by setting $\hat{F}_{\rm S}(t)=\hat{\sigma}_z$. Ref.~\onlinecite{Tuorila10} gives an example of the probe absorption spectroscopy in the case of a non-trivial quasiperiodic probe. 

\subsection{Choice of basis} \label{subsec.choiceofbasis}
As was discussed in Sec.~\ref{Floquet.subsec}, the infinite (in rank) Floquet Hamiltonian has to be truncated before its eigenproblem can be solved. The accuracy of the truncation is dependent on the choice of the atomic basis $\mathcal{B}_{\rm A}$. In the case of a strongly driven two-level system, there are two natural choices for the basis, the adiabatic and the diabatic bases (see also Ref.~\onlinecite{Silveri12}). Here, the eigenbasis of $\hat{\sigma}_z$ in (\ref{general.ham}) is called the diabatic basis. It holds the implicit assumption that the tunneling amplitude $\Delta$ is a small perturbation, $\Delta/\hbar\omega\ll 1$. Another choice for the basis is the eigenstates of the static Hamiltonian~$\hat{H}_0$. This is referred to as the adiabatic basis, which works the best when the tunneling amplitude $\Delta$ is not just a small perturbation, but of the same order as $\hbar\omega$ and $\varepsilon_0$. 

In the presence of substantial driving, one way to decide the basis preferable for the calculations is to study the LZS-dynamics~\cite{LZSM,Shevchenko09} of the driven qubit $\hat{H}(t)=\hat{H}_0+\hat{H}_{\rm S}(t)$. The probability of Landau-Zener (LZ) tunneling between the adiabatic eigenstates is given by $P_{\rm{LZ}}=\exp\left(-2\pi\Delta^2/4\hbar\omega \sqrt{A^2-\varepsilon_0^2}\right)$ for $A> \varepsilon_0$, otherwise $P_{\rm{LZ}}$ is  small. If $P_{\rm{LZ}}$ is small, the adiabatic basis is the natural choice for the basis in quasienergy calculations. In the opposite case where $P_{\rm{LZ}}\sim 1$, the diabatic basis states are closer to the eigenstates of the Floquet Hamiltonian, and thus appropriate 
for quasienergy calculations.

In the following analytic calculation of the quasienergy structures, we use the adiabatic basis when $A<\varepsilon_0$ and the diabatic basis otherwise. After solving the quasienergies, we consider the probe induced transitions between quasienergy states in the diabatic basis using the RWA. It is important to note that whereas the approximate results are basis dependent, all exact results (such as the numerical quasienergies) are not. Nevertheless, the size of the truncated Floquet Hamiltonian required for accurate results can have strong dependence on the chosen atomic basis.

%--------------------------------%
\subsection{Quasienergy states} \label{subsec.Quasienergies}
We neglect the probe and dissipation, and consider only the strongly driven qubit $\hat{H}(t)=\hat{H}_0+\hat{H}_{\rm S}(t)$. We do a transformation into a non-uniformly rotating frame with $\hat{H}'=\hat{U}^\dagger \hat{H} U+\ii \hbar (\partial_t \hat{U}^\dagger) \hat{U}$, where $\hat{U}(t)$ is a unitary time-dependent rotation~\cite{Oliver05}
\begin{equation}
  \hat{U}(t)=\exp\left(-\ii \frac{A}{2\hbar\omega}\sin(\omega t) \hat{\sigma}_z\right). \label{time-dep.u}
\end{equation}
The operation removes  the strong drive in the $z$ direction at the expense of generating  in the $x$ direction all harmonics $n\omega$ with relative weights $\Delta J_n(A/\hbar\omega)/2$. According to Sec.~\ref{Floquet.subsec}, all $\tau$-periodic entities are time-independent in the Sambe space and can be expressed in the matrix notation~\cite{Son09, Hausinger10}. 
%--------------------------------%
\begin{figure*}
\includegraphics[width=0.95\linewidth]{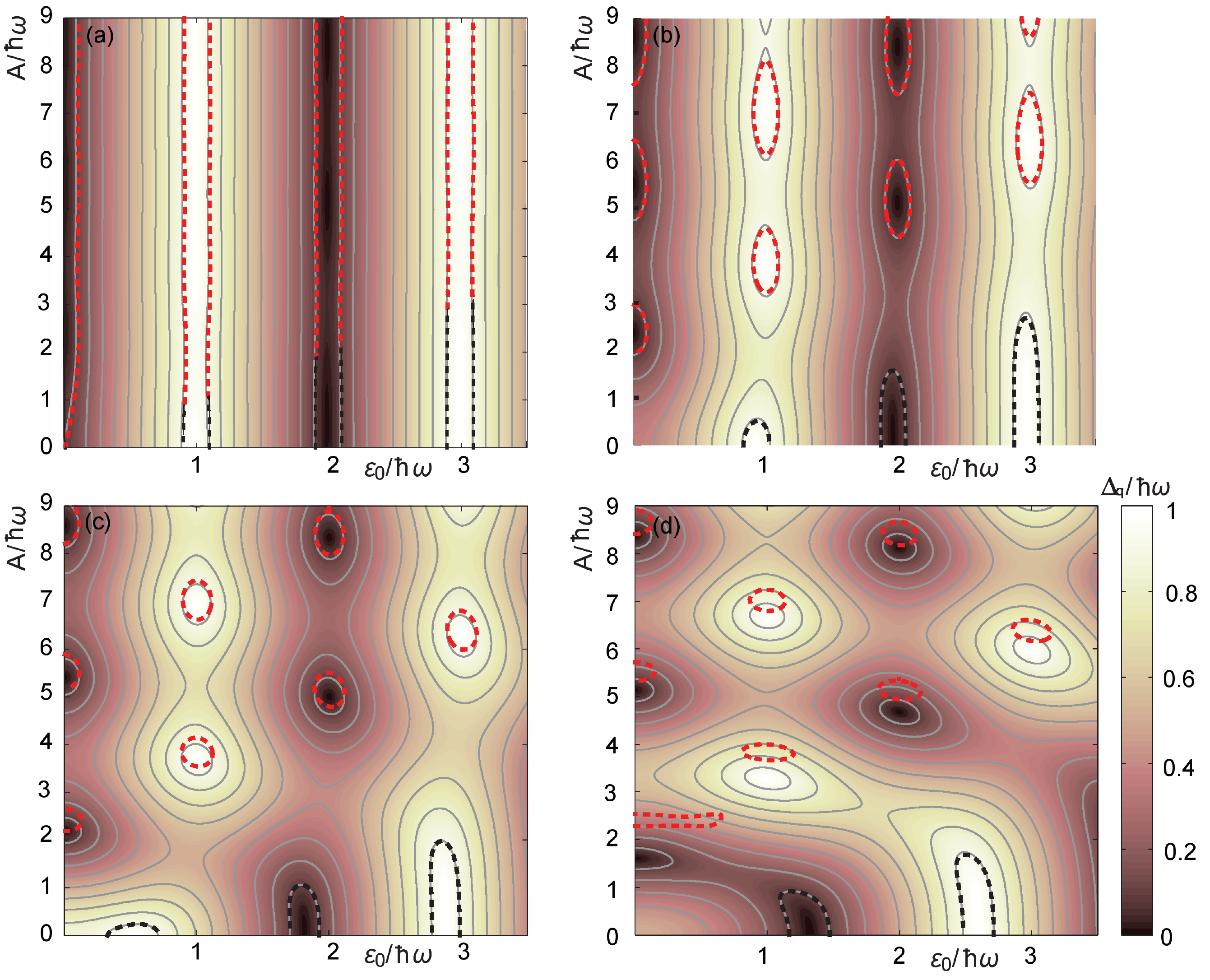}
\caption{\label{quasienergies.fig} Landscapes of the quasienergy $\Delta_{\rm q}$ in the $\varepsilon_0-A$ plane with different values for the tunnel amplitude: (a) $\Delta/\hbar\omega=0.10$; (b) $\Delta/\hbar\omega=0.37$ corresponding to an experimental realization~\cite{Wilson07}; (c) $\Delta/\hbar\omega=0.84$ corresponding to another experimental realization~\cite{Izmalkov08}; (d) $\Delta/\hbar\omega=1.50$. The energy scale is the same in all panels.  For the contour lines, we use the following color coding: numerical (solid gray), analytic in the diabatic basis Eq.~\eqref{diab_quasi.eq} (dashed red),  and analytic in adiabatic basis, that is, the adiabatic version of Eq.~\eqref{diab_quasi.eq}  (dashed black). The numerical contour lines are spaced by $0.10\, \hbar\omega$, except for two contours in the panel (b), where the comparison between the numerical and analytical contour lines is done with $\Delta_{\rm q}/\hbar\omega=0.092$ and  $\Delta_{\rm q}/\hbar\omega=0.918$  since they correspond to the values of the experimentally measured weak probe resonances~\cite{Wilson07}. }
\end{figure*}
%--------------------------------%

A resonance between the strong drive and the qubit is seen in the Sambe space as a pair of states that are nearly degenerate. Here we assume that the contribution of the non-resonant states to the resonant coupling is small. Thus, one can rely on the RWA and ignore all but the resonant states and the direct coupling between them.  We choose one pair of the resonant states, resulting in
\begin{equation}
  \hat{H}_{\rm RWA}=\begin{pmatrix} \frac{\varepsilon_0}{2} & \frac{\Delta}{2} J_n \left(\frac{A}{\hbar\omega}\right) \\  \frac{\Delta}{2} J_n \left(\frac{A}{\hbar\omega}\right)& -\frac{\varepsilon_0}{2}+n\hbar\omega \end{pmatrix}. \label{RWA.ham}
\end{equation}
There is an infinite amount of other similar pairs that are just copies of Eq.~\eqref{RWA.ham} shifted in energy due to the periodicity of the Floquet matrix $\matr{H}_{\rm F}$. 

The diagonalization of $\hat{H}_{\rm RWA}$ produces the quasienergy difference $\Delta^{\rm RWA}_{\rm q}$  
\begin{equation}
 \hat{H}_{\rm RWA}=\frac{\hat{\sigma}_z}{2}\Delta^{\rm RWA}_{\rm q}
 =\frac{\hat{\sigma}_z}{2} \sqrt{(\varepsilon_0-n\hbar\omega)^2+\Delta^2 J^2_n \left(\frac{A}{\hbar\omega}\right)} \label{quasienergy.rwa}.
\end{equation}
The diabatic-basis RWA is accurate if $\Delta/\hbar\omega\ll 1$. By following the generalized van Vleck perturbation theory~\cite{Certain70, Son09, Hausinger10}, the RWA result \eqref{quasienergy.rwa} can be corrected with higher-order terms in the perturbation parameter $\Delta/\hbar\omega$. The second order correction~\cite{Autler55,Son09,Tuorila10} affects the locations of the strong driving resonances: $\varepsilon_0=n\hbar\omega-\delta$. We call it the $\delta$ shift. In the diabatic basis, the explicit expression for $\delta$ shift is
\begin{equation}
  \delta_{{\rm d}}=2\mathop{\sum_{k=-\infty}^{\infty}}_{k\neq n}  \frac{\left[\frac{\Delta}{2} J_k \left({A}/{\hbar\omega}\right)\right]^2}{\varepsilon_0+k\hbar\omega}. \label{delta.eq}
\end{equation}
The corrected quasienergy splitting is then
\begin{equation}
  \Delta^{{\rm RWA}+\delta}_{\rm q}=\sqrt{(\varepsilon_0+\delta_{\rm d}-n\hbar\omega)^2+\Delta^2 J^2_n \left(\frac{A}{\hbar\omega}\right)}. ~\label{diab_quasi.eq}
\end{equation}
 In the diabatic basis, the $\delta_{\rm d}$ shift is the most important at small amplitude $A$ and at small $n$. This implies that the $\delta_{\rm d}$-shift vanishes with moderate driving amplitudes~\cite{Son09}, that is, when the diabatic basis is the most natural choice for the basis.  

In the adiabatic basis, one first diagonalizes $\hat{H}_0$ and then transforms $\hat{H}(t)$ to the non-uniformly rotating frame with a time-dependent transformation analogous to~\eqref{time-dep.u}. The resulting Floquet matrix has exactly the same structure as in the diabatic basis, but the diabatic diagonal energy $\varepsilon_0$ is replaced by $\hbar\omega_0=\sqrt{\varepsilon_0^2+\Delta^2}$ and diabatic coupling strength 
\begin{equation}
 \Delta J_n(A/\hbar\omega)/2 \rightarrow \frac{n\hbar\omega\Delta}{2\varepsilon_0} J_n\left(\frac{A}{\hbar\omega} \frac{\varepsilon_0}{\hbar\omega_0}\right). \label{Omega_k}
\end{equation}
The adiabatic resonance condition can then be  written as $\hbar\omega_0=n\hbar\omega-\delta_{\rm a}$, where the adiabatic $\delta_{\rm a}$ shift is calculated with the formula~\eqref{delta.eq}, but using the adiabatic coupling strengths~\eqref{Omega_k} and the diagonal energies $\hbar\omega_0$. 

In Fig.~\ref{quasienergies.fig}, we have shown the comparison of the numerically and analytically calculated quasienergy landscapes in the $\varepsilon_0-A$ plane. The adiabatic basis (black dashed) is applied when $A<\varepsilon_0$ and the diabatic basis (red dashed) otherwise. The analytic quasienergies agree well with the corresponding numerical ones when the effects of tunnel amplitude can be handled with the perturbation theory, cf. Fig.~\ref{quasienergies.fig}(a)-(b). But, the generalized van Vleck perturbation theory becomes insufficient~\cite{Hausinger10} if the fraction $A/\Delta$ becomes large enough, and simultaneously $\Delta/\hbar\omega>1$. In this limit, the calculation of the quasienergies is necessarily numerical. The breakdown of the analytical approach is demonstrated in Fig.~\ref{quasienergies.fig}(c)-(d).

\subsection{Weak probe transitions}
%--------------------------------% 
\begin{figure}%[tbp!]
\includegraphics[width=1.0\columnwidth]{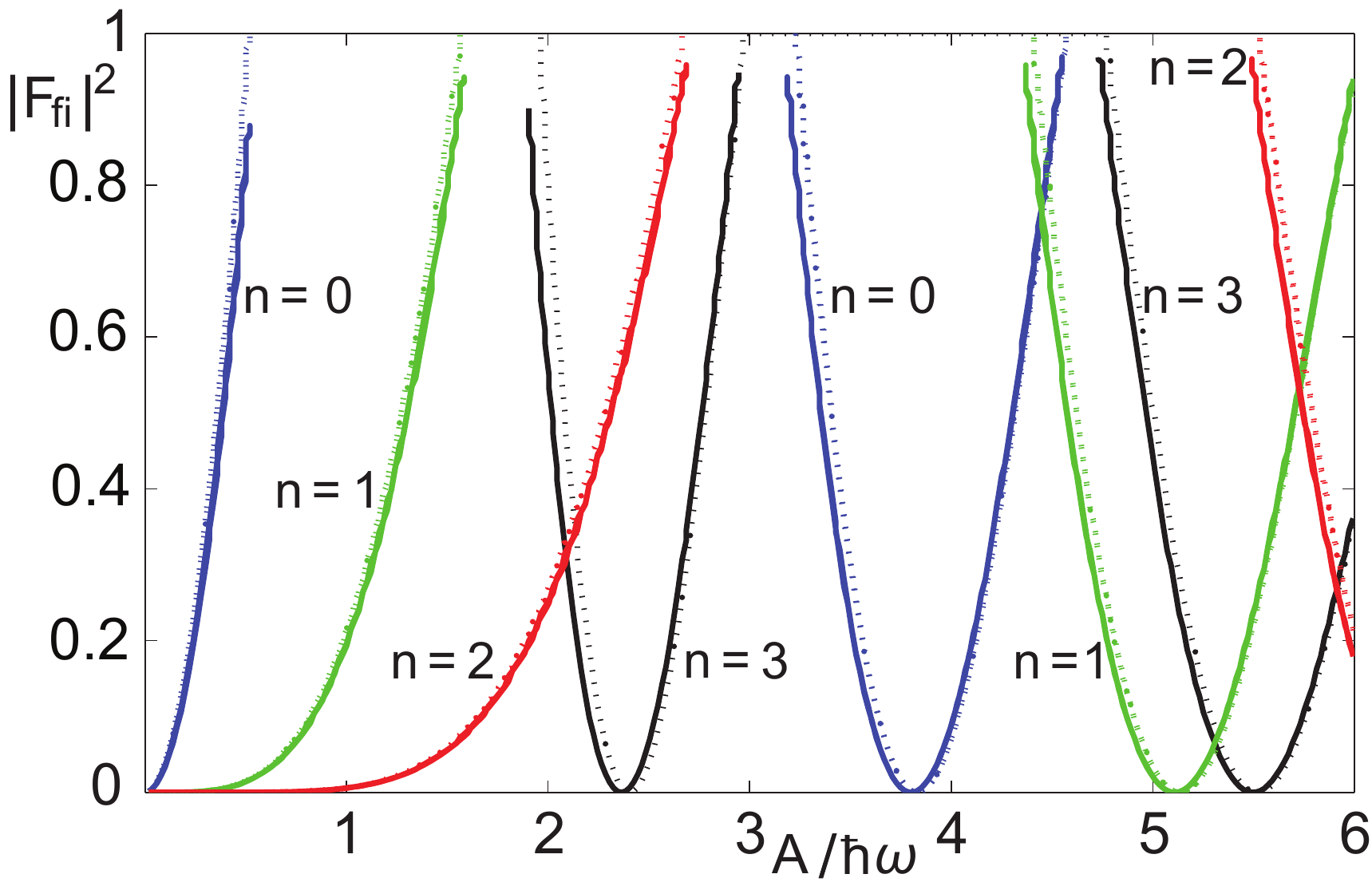}
\caption{Comparison between the numerical (solid line) and RWA analytical (dotted line) transition amplitudes $|F_{fi}|^2$~\eqref{Analytical.pfi}, calculated with the parameters of Fig.~\ref{quasienergies.fig}(b). The transition amplitudes are picked by following the corresponding resonance conditions~\eqref{res.cond.eq}, i.e., from the parametrized line $\varepsilon_0(A)$ where $\Delta_{\rm q}=0.092\hbar\omega$ or $\Delta_{\rm q}=0.918\hbar\omega$. The curves are labeled by the index $n$ in Eq.~\eqref{RWA.ham}.
\label{anal_comp.fig}}
\end{figure}
%--------------------------------%
We now discuss the probe resonance condition and the probe transition elements $F_{fi}$~\eqref{Numerical.pfi} in terms of the diabatic basis and the RWA. This kind of treatment is adequate for the essential physical insight. We follow the same procedure as in calculating quasienergies. The transformation $\hat{U}(t)$~\eqref{time-dep.u} does not change the probe part of Hamiltonian~\eqref{general.ham}. In the subsequent transformation to the Sambe space, the $\tau$-periodic part of the weak probe obtains the form $\hat{F}_{\rm S}=\hat{\sigma}_z\otimes\matr{\mathbbm{I}}$, where $\matr{\mathbbm{I}}$ denotes the infinite-dimensional identity matrix. 
%--------------------------------% 
\begin{figure*}%[tbp!]
\includegraphics[width=0.95\linewidth]{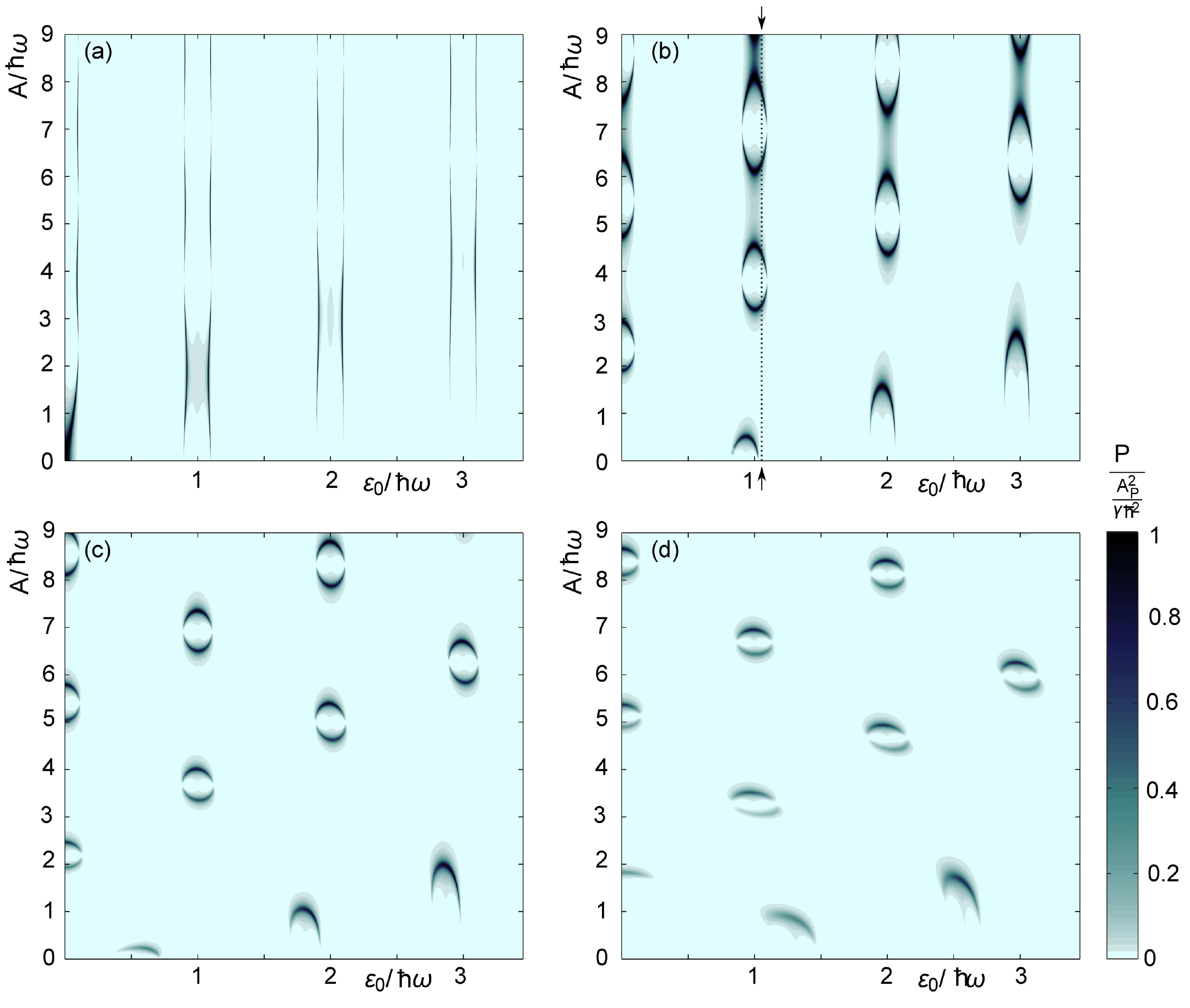}
\caption{ \label{lines.fig}The transition rate $\mathcal{P}$~\eqref{goldenrule.sum.eq} of the strongly driven and weakly probed qubit presented as a gray-scale plot in the $\varepsilon_0-A$ plane. The parameters are the same as in the corresponding panels in Fig.~\ref{quasienergies.fig}. In addition to those, the parameter $\kappa$ describing the Ohmic spectral density is chosen so that $\gamma/\omega=0.016$  in the absence of driving and detuning in Eq.~\eqref{gamma.fbm} and $\beta=\hbar\omega/kT=2.24$. The discontinuities  of the lines are a consequence of the roots of $J_n(A/\hbar\omega)$ related to the coherent destruction of the tunneling. The $\mathcal{P}$ versus $A$ plot along the vertical dashed line in panel (b) is  shown in  Fig.~\ref{lines_comp.fig}.}
\end{figure*}
%--------------------------------% 

As the strongly driven part of the total Hamiltonian is truncated into a two-level system~\eqref{RWA.ham}, it is reasonable to make the same reduction for the weak probe part. The $\tau$-periodic part of the weak probe becomes simply $\hat{F}_{\rm S}=\hat{\sigma}_z$, operating between the resonant basis states of the RWA Hamiltonian~\eqref{RWA.ham}. In the diagonalization of the strongly driven part of the Hamiltonian, the perturbation matrix  gets a non-diagonal form
\begin{align}
  \hat{F}_{\rm S}&=\frac{n\hbar\omega-\varepsilon_0}{\Delta_q^{\rm RWA}}\hat{\sigma}_z +\frac{\Delta J_n \left(\frac{A}{\hbar\omega}\right)}{\Delta_{\rm q}^{\rm RWA}}\hat{\sigma}_x, \label{RWA.probe.ham}
\end{align}
expressed directly in the basis of quasienergy states $\ket{u_{+}}$ and $\ket{u_{-}}$ with the energy splitting $\Delta_{\rm q}^{\rm RWA}$~\eqref{quasienergy.rwa}. The longitudinal weak probe itself would not induce transitions between the non-driven diabatic eigenstates, but the rotation to qubit eigenbasis ($\propto\Delta$) and the dressing of the strong drive [$\propto J_n (A/\hbar\omega)$] have such a effect that probe transitions become possible. This is formally seen as the non-zero transverse $\hat{\sigma}_x$ term in Eq.~\eqref{RWA.probe.ham}. 

In the general case, the resonance condition for the probe transition is
\begin{equation}
  \Delta_{\rm q}=\begin{cases} \hbar\omega_{\rm P}-k \hbar\omega, \\  (k+1) \hbar\omega-\hbar\omega_{\rm P},\end{cases} \label{res.cond.eq}
\end{equation}
where $k=0,1,2,\ldots$ is chosen so that $k\omega < \omega_{\rm P} < (k+1)\omega$. The quasienergy difference $\Delta_{\rm q}$ is defined as the difference between two consecutive quasienergy levels. We consider now the case $k=0$ (cf. Fig~\ref{TemporalQuasienergy.fig}). Thus, the weak probe transitions are possible when the quasienergy difference $\Delta_{\rm q}=\hbar\omega_{\rm P}$ or $\Delta_{\rm q}=\hbar\omega-\hbar\omega_{\rm P}$, and the corresponding transition matrix element is non-zero. In Fig.~\ref{quasienergies.fig}, the resonance condition is shown as highlighted contour lines (dashed lines). If the tunneling amplitude $\Delta$ is so large that the minimum quasienergy difference $|\Delta J_n\left({A}/{\hbar\omega}\right)|$ is larger than the probe energy $\hbar\omega_{\rm P}$, there are no resonances. In Fig.~\ref{quasienergies.fig}(a), the tunneling amplitude $\Delta$ is small and the resonances are continuous lines in vertical direction, but as the value of $\Delta$ is increased the resonances curve and close, cf. Fig.~\ref{quasienergies.fig}(b)-(d). 

The matrix element~\eqref{Numerical.pfi} for the weak probe transition between the quasienergy states $\ket{u_{+}}$ and $\ket{u_{-}}$ is directly the non-diagonal element in~\eqref{RWA.probe.ham}
\begin{align}
  |F_{fi}|^2=\left|\inner{u_{+}}{\hat{F}_{\rm S}}{u_{-}}\right|^2=\frac{\Delta^2 J^2_n \left(\frac{A}{\hbar\omega}\right)}{\left(\Delta^{\rm RWA}_{\rm q}\right)^2}.  \label{Analytical.pfi}
\end{align}
We are interested in the transition element $|F_{fi}|^2$ when the weak probe is (nearly) resonant, that is, $\Delta^{\rm RWA}_{\rm q}\approx\hbar\omega_{\rm P}$. Thus, the transition amplitude depends only on the coupling strength $\Delta J_n(A/\hbar\omega)$ of the two uncoupled energy levels in Eq. \eqref{RWA.ham}. The comparison between the numerical (solid) and analytical (dotted) transition amplitudes is shown in Fig.~\ref{anal_comp.fig}. It is calculated by following the weak probe resonances~\eqref{res.cond.eq}. The agreement between the numerical and analytical results is good by taking into account that the chosen parameters are close to the validity boundary of the RWA.

To calculate the transition rate $\mathcal{P}$~\eqref{goldenrule.sum.eq}, in addition to the quasienergies and the quasienergy states, one needs the dephasing rate $\gamma_{ij}=\gamma$ and the populations $p_i$ of the quasienergy states. We estimate them by following Refs.~\onlinecite{Hausinger10, Wilson10} that apply the Floquet-Born-Markov-formalism~\cite{OpenQuantumSystems, GrifoniHnggi98, Wilson10, Hausinger10, Kohler97, Goorden, Hone09}, which successfully merges the Floquet method and detailed coupling to the environment. First, one constructs the master equation for the strongly driven qubit coupled to the environment through the $\hat{\sigma}_z$ operator, i.e., via the matrix elements $X_{\alpha\beta n}=\inner{u_{\alpha,0}}{\hat{\sigma}_z\otimes \matr{\mathbbm{I}}}{u_{\beta,n}}$. The quasienergy states $\ket{u_{\beta,n}}$ are employed to calculate the above matrix elements $X_{\alpha\beta n}$. Here this is done numerically, but it can also be done analytically with the RWA or with the second order Van Vleck-correction, within their validity ranges~\cite{Hausinger10}. The environment is modeled with a continuum of harmonic oscillators, i.e., a thermal bath characterized with $N_{\alpha\beta n}=\frac12G_{\alpha\beta n}\left\{\coth\left[\hbar(\epsilon_\alpha-\epsilon_\beta+n\omega)\hbar\omega/2kT\right] -1\right\}$ and the Ohmic spectral density $G_{\alpha\beta n}=G(\varepsilon_\alpha-\varepsilon_\beta+n\omega)=\kappa(\varepsilon_\alpha-\varepsilon_\beta+n\omega)$. Finally, the coefficients in the master equation are averaged over the period of the strong drive, in order to bring them into time-independent form~\cite{Kohler97,Hone09} (secular approximation, moderate rotating wave approximation). The result is analytically solvable in the steady state limit~\cite{Hausinger10}, from which the dephasing rate $\gamma$ and the population $p_{-}$ ($p_{+}=1-p_{-}$) are derived
\begin{align}
&\gamma=\pi \sum_{n=-\infty}^\infty(2N_{-+n}+G_{-+n})X_{-+n}^2+4N_{--n}X_{--n}^2, \label{gamma.fbm}\\
  & p_{-}=\frac{\sum_{n=-\infty}^\infty N_{-+n}X^2_{-+n}}{\sum_{n=-\infty}^\infty (2N_{-+n}+G_{-+n})X_{-+n}^2}. \label{quasi_pop.eq}
\end{align}
The numerically calculated transition rates $\mathcal{P}$~\eqref{goldenrule.sum.eq} are shown in Fig.~\ref{lines.fig} in the $\varepsilon_0-A$ plane.

The total line-shape~\eqref{goldenrule.sum.eq} encodes the information on the quasienergy structure at the locations of the resonances and on the transition amplitudes in the magnitudes of the resonances. By comparing the line-shapes in Fig.~\ref{lines.fig} with the quasienergy structure of Fig.~\ref{quasienergies.fig}, one observes the faithful mapping of the energy landscape.  The maximum value of the transition element $|F_{fi}|^2$~\eqref{Analytical.pfi} depends on the tunneling amplitude $\Delta$. If $\Delta$  is large enough, the maximum is reached when $\Delta J_n(A/\hbar\omega)=\hbar\omega_{\rm P}$. The transition element cannot obtain larger values since then the resonance condition is not anymore valid, see Eq.~\eqref{quasienergy.rwa} and Fig.~\ref{lines.fig}(b). With smaller tunneling amplitude $\Delta$, the maximum of the $|F_{fi}|^2$ is directly set by the maximum of $J_n(A/\hbar\omega)$, see Fig.~\ref{lines.fig}(a). The weak probe signal vanishes at the zeros of the $J_n(A/\hbar\omega)$, which are related to the coherent destruction of tunneling~\cite{Grossmann91}.  This is seen in Fig.~\ref{lines.fig} as discontinuous resonance lines, although the underlying quasienergy resonance conditions are continuous lines (a) or closed curves (b)-(d).

\subsection{Relation to the spectrum of the probe field}
%--------------------------------% 
\begin{figure}%[tbp!] 
\includegraphics[width=1.0\columnwidth]{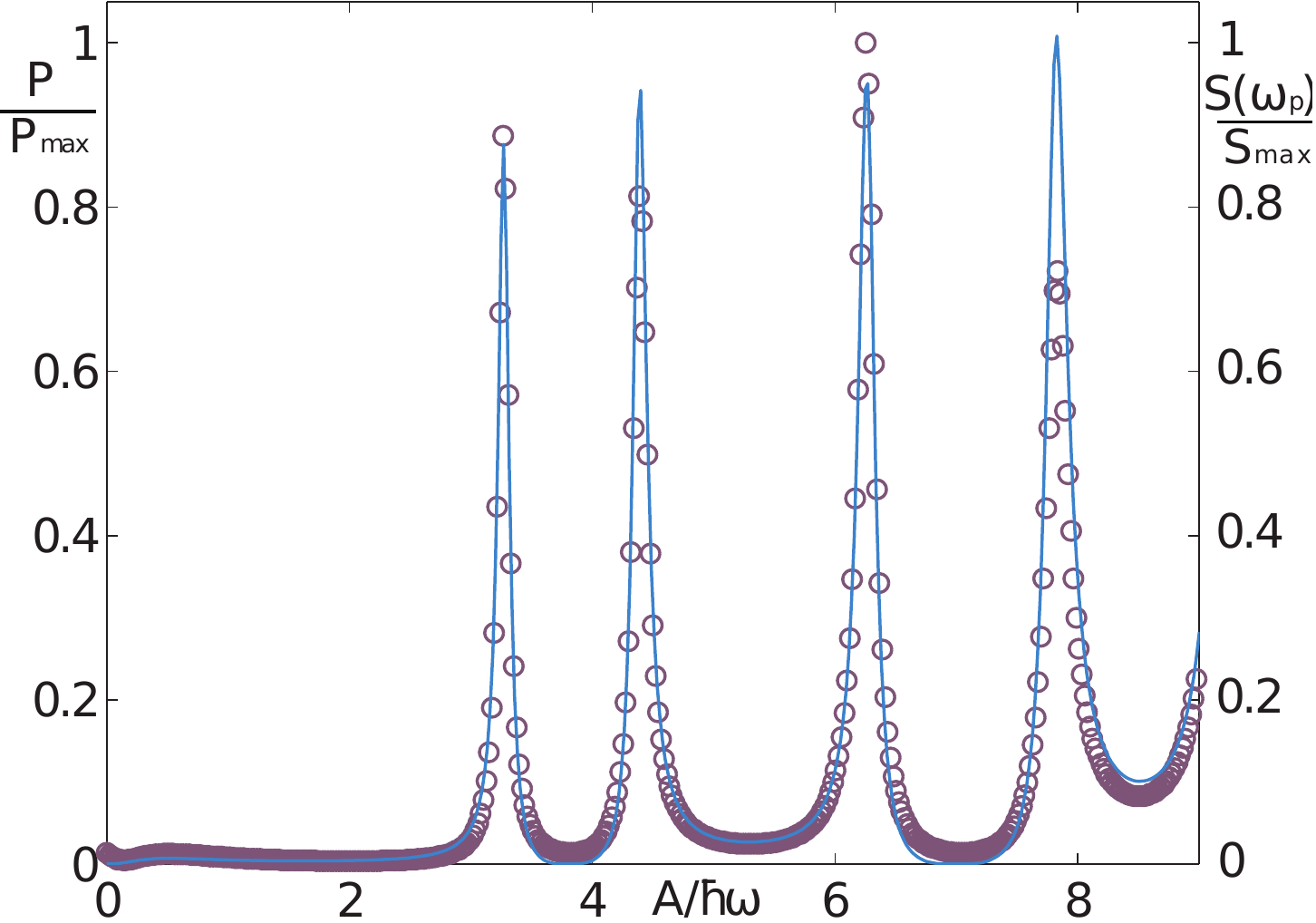}
\caption{\label{lines_comp.fig} The transition rate $\mathcal{P}$ of the strongly driven and weakly probed qubit and the corresponding spectrum $S(\omega_{\rm P})$, both scaled with their maximum value. The parameters are $\varepsilon_0/\hbar\omega=1.05$, $\Delta/\hbar\omega=0.37$,and $\omega_{\rm P}/\omega=0.092$. The parameter $\kappa$ describing the Ohmic spectral density is chosen so that $\gamma/\omega=0.016$ in the absence of driving and detuning in Eq.~\eqref{gamma.fbm} and $\beta=\hbar\omega/kT=2.24$. The transition rate $\mathcal{P}$ is calculated with the numerical Floquet method [solid line, vertical projection from Fig.~\ref{lines.fig}(b)], which is contrasted with the spectrum $S(\omega_{\rm P})$~\eqref{eq:spec_corr_zz} (circles) of the driven qubit ($1/T_2=\gamma$ and $1/T_1=\gamma/2$) at the weak probe frequency.}
\end{figure}
%--------------------------------% 
In the case of the two-level system \eqref{general.ham}, the spectrum as a function of the correlator (\ref{inoutput.spec.eq}) takes the form
\begin{equation}
  S(\omega_{\rm P})=\frac{A_{\rm P}^2}{16\hbar^2}\int_{-\infty}^\infty \expe{\hat{\sigma}_z(t)\hat{\sigma}_z (0)}\ee^{\ii \omega_{\rm P} t} \D{t}. \label{eq:spec_corr_zz} 
\end{equation}
Noteworthily, this spectrum is not the one commonly calculated from the transverse correlator $\expe{\hat{\sigma}_-(\tau)\hat{\sigma}_+ (0)}$, natural to the atomic systems coupling to the environment through the (transverse) dipole moment. In Fig.~\ref{lines_comp.fig}, we have compared the line-shape $\mathcal{P}$ calculated by using numerically implemented Floquet method (solid line) and weak probe response (circles) $S(\omega_{\rm P})$~\eqref{eq:spec_corr_zz}, obtained by solving the steady state master equation. The correlation function approach (circles) agrees very well with the transition rate calculated with the numerical Floquet method (solid), which further validates the method of the probe spectroscopy of quasienergies. The slight differences between the two methods can be traced back to the different approximations concerning relaxation and dephasing. In contrast to the detailed Floquet-Born-Markov-formalism, the master equation of the qubit corresponding to~\eqref{eq:spec_corr_zz} includes simply the standard relaxation and dephasing, with rates $1/T_1$ and $1/T_2$, respectively.
 
%--------------------------------%
%--------------------------------% 
\subsection{Comparison with experiments}\label{Exp_sec}
We have also interpreted two recent experiments in terms of probe absorption of quasienergy states. The experiment by Wilson et al.~\cite{Wilson07} uses a  Cooper-pair box and the experiment by  Izmalkov et al.~\cite{Izmalkov08} uses a flux qubit, but both can be described by the Hamiltonian in Eq.~\eqref{general.ham}.

\begin{figure}
\centering
\includegraphics[width=1.0\columnwidth]{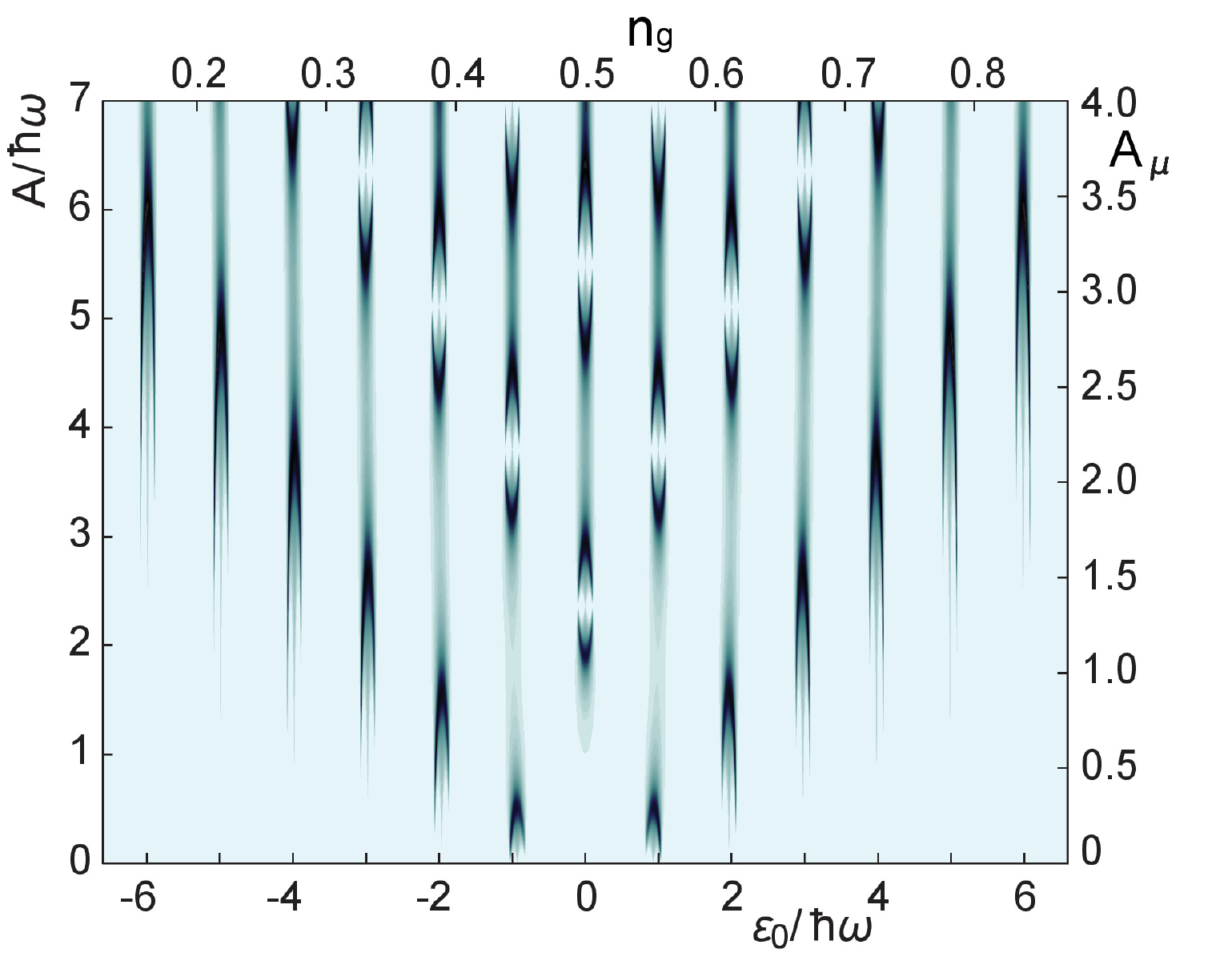}  
\caption{\label{Wilson.compare}The probe absorption $\mathcal{P}$ calculated as a function level spacing $\varepsilon_0$ and driving amplitude $A$. The parameters are $\omega/2\pi=7.0$~GHz, $\Delta/\hbar\omega=0.37$, $\omega_{\rm P}/\omega=0.092$, and $T=150$~mK. The parameter $\kappa$ describing the Ohmic spectral density is chosen so that $\gamma/\omega=0.045$ in the absence of driving and detuning in Eq.~\eqref{gamma.fbm}, corresponding the experimental estimate for qubit dephasing.   This plot should be compared with the experimental plot in Ref.~\onlinecite{Wilson07}. For the comparison we have given the axis scales also using the units of this reference.}
\end{figure}

Figure~\ref{Wilson.compare} shows the calculated probe absorption corresponding to the experiment of Wilson et al\cite{Wilson07}. The parameters are the same as given in Ref.~\onlinecite{Wilson07} except that we have not included the extra broadening caused by low-frequency fluctuations in the gate charge $n_{\rm g}$. We have used the same parameters also in Figs.~\ref{quasienergies.fig}(b) and \ref{lines.fig}(b), except that  the line-width $\gamma$ is almost three times larger than in Fig.~\ref{lines.fig}(b). The resonances in Fig.~\ref{Wilson.compare} still have the same characteristic features as in Fig.~\ref{lines.fig}(b), but they are not as clear because of the larger line width. Figure~\ref{Wilson.compare} should be compared with the experimental plot  in Ref.~\onlinecite{Wilson07} which, however, has the extra broadening that wipes out some of the features. In the same reference, the experimental data was successfully compared with theory by using RWA, which is still sufficient at the parameter values of the experiment (see Fig.~\ref{quasienergies.fig}).
%--------------------------------% 
%--------------------------------%
 
\begin{figure}
\centering
 \includegraphics[width=1.0\columnwidth]{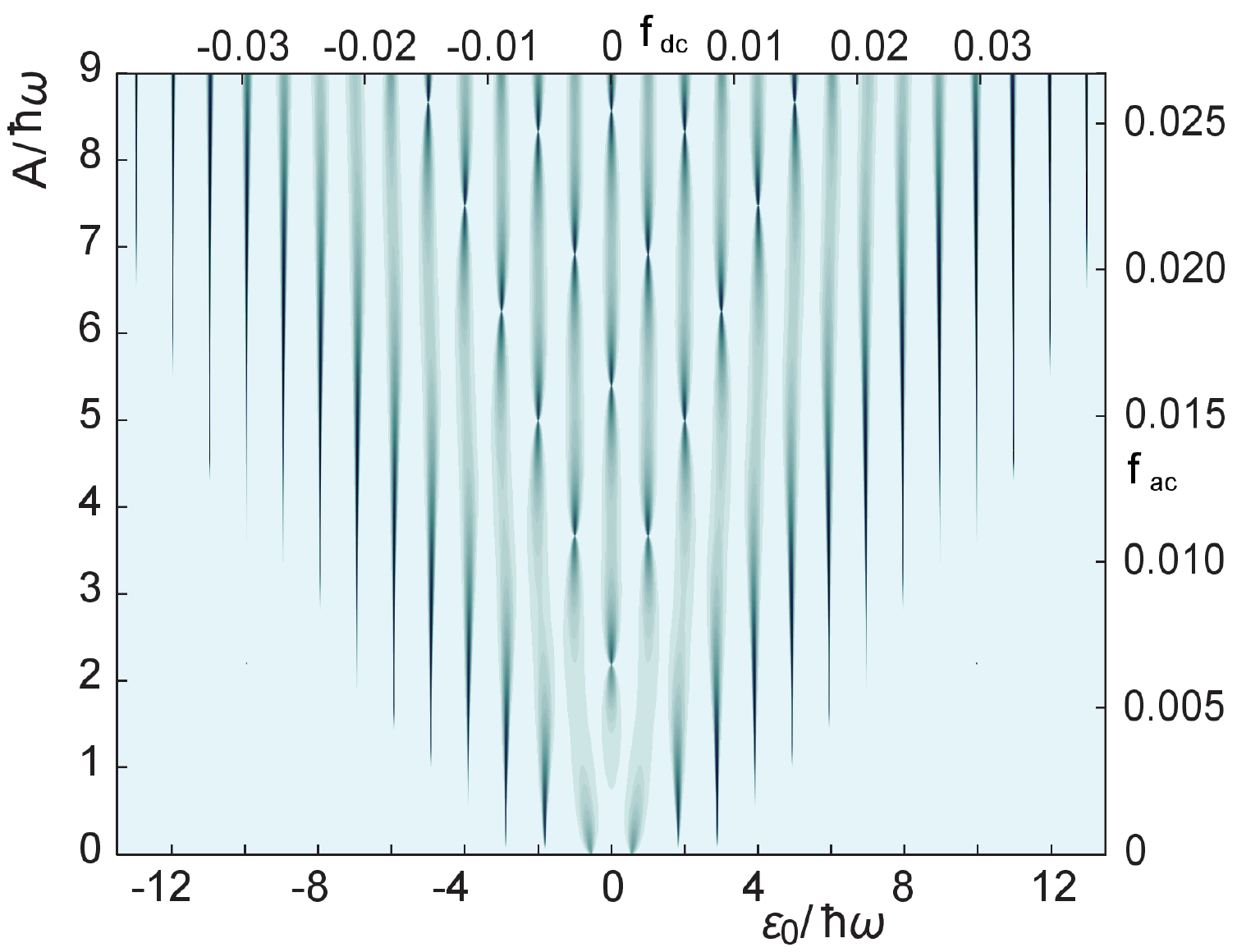}
\caption{\label{Shevchenko_line.fig}The probe absorption $\mathcal{P}$ calculated as a function level spacing $\varepsilon_0$ and driving amplitude $A$. The parameters are $\omega/2\pi=4.15$~GHz,  $\Delta/\hbar\omega=0.84$,  $\omega_{\rm P}/\omega=0.005$, and $T=70$~mK. The parameter $\kappa$ describing the Ohmic spectral density is chosen so that $\gamma/\omega=0.17$ in the absence of driving and detuning in Eq.~\eqref{gamma.fbm}, corresponding to the experimental estimate for qubit dephasing.  This plot should be compared with the experimental plot in Ref.~\onlinecite{Izmalkov08}. For the comparison we have given the axis scales also using the units of this reference.}
\end{figure}

Figure~\ref{Shevchenko_line.fig} shows the probe absorption calculated with the parameters corresponding to the experiment of Izmalkov et al\cite{Izmalkov08}. The same parameters are also used in Figs.~\ref{quasienergies.fig}(c) and \ref{lines.fig}(c), except that the probe frequency $\omega_{\rm P}/\omega=0.005$ is much smaller than in Fig.~\ref{lines.fig}(c). Now, the elliptical shape of the resonances is not resolved because of line broadening, but the discontinuities of the resonances remain. The $\delta_{\rm d}$ shift~\eqref{delta.eq}, which is a signature of the RWA breakdown, is clearly visible as the bending of the resonances as a function of driving amplitude $A$.  The $\delta_{\rm d}$ shift is enhanced near $\varepsilon_0=0$ and at small $n$ \cite{Son09}. The plot should be compared with the experimental plot  in Fig.~3 by Izmalkov et al. \cite{Izmalkov08}, taking into account that it is a phase plot instead of an absorption plot. Both plots reveal the same quasienergy landscape, as the resonances are visible as bluish lines and the discontinuities as yellow crosses in the phase plot. In the same reference the experimental results are interpreted as LZS-interferometry \cite{Shevchenko09} which produces oscillations of the qubit population. 
%--------------------------------% 
%--------------------------------% 

The discussed experiments reveal information about quasienergy landscape, but suffer from noise that prevents the observation of individual contour lines. We point out Ref.~\onlinecite{Tuorila10} as an example of an experiment where individual contour lines are clearly seen. Another difference in this experiment is that the modulation of the energy is non-sinusoidal, leading into genuinely quasiperiodic probe, in contrast to Eq.~\eqref{general.ham}. The Floquet analysis at the parameters of this experiment was reported in conjunction with the measurement (see Supplementary Information of Ref.~\onlinecite{Tuorila10}). 

%--------------------------------% 
%--------------------------------% 
\section{Conclusions}\label{Conc_sec}
We have presented a method to map the quasienergies of a driven quantum system by using a weak probe. We made the derivation with a general form of the probe Hamiltonian, but applied it to simple cases in order to gain physical insight. Provided that the quasienergy excitation has a long enough life time, the spectroscopy enables an accurate mapping of the quasienergy structures \cite{Tuorila10}. The results rely on first order perturbation expansion in the probe amplitude. We also suggested the generalized Floquet method as a possible way to go beyond the perturbative-probe approximation. 

The detailed discussion about the strongly driven and weakly probed qubit shows that, with certain parameter values, analytical results may be obtained for the weak probe resonances and the transition amplitudes, thus resulting both the absorption and dispersion of the probe response, i.e., the generalized probe susceptibility. Otherwise, numerical calculations are a necessity. However, relying only on proper matrix truncation and inversion, the solutions are numerically stable and simple to find. We noted that the accuracy of the analytic, and to some extent the numerical, calculation is dependent on the choice of the atomic basis. Indeed, the detailed study of the transition from the adiabatic to the diabatic behaviour would be interesting and possible by using the probe absorption spectroscopy of quasienergies.

We reinterpreted two recent experiments~\cite{Wilson07,Izmalkov08}. Although the estimated life-time of the quasienergy excitations in the referred experiments were too short to distinguish the quasienergy contours, we were able to point out features in the measured responses that stem from the underlying quasienergy landscape. 

%--------------------------------% 
\begin{acknowledgments}
 We thank Pekka Pietil\"ainen, Mikko Saarela, Mika Sillanp\"a\"a, and Pertti Hakonen for useful discussions. This work was financially supported by the Magnus Ehrnrooth Foundation, the Finnish Academy of Science and Letters (Vilho,Yrj\"o and Kalle V\"ais\"al\"a Foundation), and the Academy of Finland.
\end{acknowledgments}
%--------------------------------% 

%--------------------------------% 
%--------------------------------% 
\appendix*
\section{The generalized Floquet method}\label{App.two-mode}
The generalization of the Floquet method was developed in Ref.~\onlinecite{Chu83}. It enables the handling of bi- or polychromatic driving fields in a way similar to the monochromatic case. Here, we use the two-mode Floquet method in the analysis of the strongly driven and weakly probed qubit. We assume the bichromatic Hamiltonian defined in 
Eq.~\eqref{general.ham}. In the generalized Floquet picture, the solution of the time-dependent 
Schr\"odinger equation is given in the form
\begin{align}
  \ket{\Psi(t)}&=\ee^{-\ii \epsilon t/\hbar}\ket{u(t)} \\ \left(-\ii\hbar\frac{\D{}}{\D{t}}+\hat{H}(t)\right)\ket{u(t)}&=\epsilon\ket{u(t)}
\end{align}
where the quasienergy state $\ket{u(t)}$ is also bichromatic and the quasienergies are quasiperiodic 
$\epsilon=\epsilon_{r,n_1,n_2}=\epsilon_{r}+\hbar n_1 \omega+\hbar n_2 \omega_{\rm P}.$
%--------------------------------% 
\begin{figure*}
\includegraphics[width=1.0\linewidth]{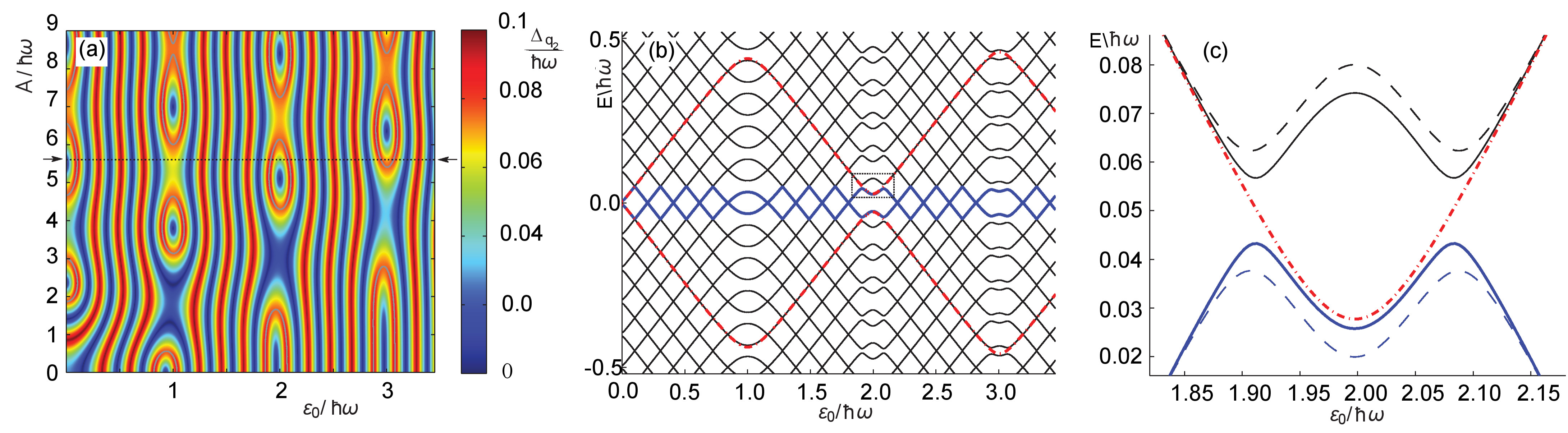}
\caption{\label{quasienergy_2_mode.fig} Quasienergy in the generalized Floquet method. (a) Quasienergy landscape in the $\varepsilon_0-A$ plane with the parameters $\Delta/\hbar\omega=0.37$, $\omega_{\rm P}/\omega=0.10$, and $A_{\rm P}/{\hbar \omega}=0.20$; The gray line shows the weak probe resonance condition~\eqref{res.cond.eq} deduced from the corresponding single-mode quasienergy landscape [Fig.~\ref{quasienergies.fig}(b)]. (b) Projection of (a) at $A/\hbar\omega=5.6$, denoted with arrows and dashed line. The quasienergy $\epsilon_{r,n,m}$ (black) is periodic so that $\epsilon_{r}$ (blue) is shifted by $n\hbar\omega+m\hbar\omega_{\rm P}$, where $n,m=0,\pm 1, \pm 2, \ldots$. The red dash-dotted line shows the corresponding quasienergy calculated with the single-mode Floquet-method.  (c) Comparison of the two-mode quasienergies with the probe amplitude $A_{\rm P}/{\hbar \omega}=0.20$ (solid) and $0.40$ (dashed). Magnified view of the box in panel (b).}
\end{figure*}
%--------------------------------% 

To take advantage of the periodicity, we express the Hamiltonian~\eqref{general.ham} and the state $\ket{u(t)}$ using a 'double' Fourier series representation:
\begin{align}
  \hat{H}(t)&=\mathop{\sum_{n_1=-\infty}^{\infty}}_{n_2=-\infty} \sum_{\sigma,\sigma'} \ee^{\ii (n_1 \omega+n_2 \omega_{\rm P})t} h^{(n_1),(n_2)}_{\sigma\sigma'} \ket{\sigma}\bra{\sigma'}
\end{align}
\begin{align}
\ket{u(t)}&=\mathop{\sum_{n_1=-\infty}^{\infty}}_{n_2=-\infty}\sum_{\sigma} \ee^{\ii (n_1 \omega+ n_2 \omega_{\rm P}) t} c^{(n_1),(n_2)}_\sigma \ket{\sigma}.
\end{align}
Similar to the case of the single-mode Floquet method [see Eq.~\eqref{FloquetMatrix.eq}], we get a time-independent eigenvalue equation  
\begin{align}
  \matr{H}_{\rm F_2}\ket{u}=\epsilon \ket{u}. \label{2-mode.floquet.eq}
\end{align}
The Hamiltonian~\eqref{general.ham}
can be expressed in terms of the sub matrices $\matr{H}^{[0]}=\frac12\left(\varepsilon_0 \hat{\sigma}_z+\Delta \hat{\sigma}_x\right)$, $\matr{H}^{[\pm 1]}=\frac{A}{4}\hat{\sigma}_z$, and $\matr{B}^{[\pm 1, 0]}=\frac{A_{\rm P}}{4}\hat{\sigma}_z$:
\begin{align}
\hat{H}(t)=\matr{H}^{[0]}+&\matr{H}^{[\pm1]} \left( \ee^{\ii\omega t}+\ee^{-\ii\omega t}\right)\notag\\&+\matr{B}^{[\pm1, 0]} \left( \ee^{\ii\omega_{\rm P} t}+\ee^{-\ii\omega_{\rm P} t}\right). 
\end{align}
The two-mode Floquet matrix $\matr{H}_{\rm F_2}$ of the Hamiltonian is given as an infinite dimensional matrix~\cite{Chu83}
\begin{equation}
\matr{H}_{\rm F_2}=\begin{pmatrix}
\ddots& &\vdots&& \iddots\\
&\matr{H}_{\rm F}-\matr{\mathbbm{I}}\hbar\omega_{\rm P} &{\matr{B}}^{[1]} &\matr{0}&\\
\cdots&{\matr{B}}^{[-1]} & \matr{H}_{\rm F} & {\matr{B}}^{[1]} &\cdots\\
&\matr{0} & {\matr{B}}^{[-1]} & \matr{H}_{\rm F}+\matr{\mathbbm{I}}\hbar\omega_{\rm P} &\\
\iddots&&\vdots&&\ddots
\end{pmatrix}. \label{hf2.eq}
\end{equation}
All the entries in $\matr{H}_{\rm F_2}$ are matrices of infinite rank. The single-mode Floquet matrix $\matr{H}_{\rm F}$ is on the diagonal and it has the familiar form
\begin{equation}
\matr{H}_{\rm F} =\begin{pmatrix}
\ddots& &\vdots&& \iddots\\
&\matr{H}^{[0]}-\matr{\mathbbm{I}}\hbar\omega &{\matr{H}}^{[1]} &\matr{0} &\\
\cdots&{\matr{H}}^{[-1]} & \matr{H}^{[0]} & {\matr{H}}^{[1]} &\cdots\\
&\matr{0} & {\matr{H}}^{[-1]} & \matr{H}^{[0]}+\matr{\mathbbm{I}}\hbar\omega &\\
\iddots& &\vdots&& \ddots
\end{pmatrix}.
\end{equation}
In $\matr{H}_{\rm F_2}$~\eqref{hf2.eq}, the $k\hbar \omega_{\rm P}$-shifted single-mode entries $\matr{H}_{\rm F}+\matr{\mathbbm{I}}k\hbar\omega_{\rm P}$ are coupled by infinite-rank coupling matrices $\matr{B}^{[n]}$, defined as
\begin{equation}
\matr{B}^{[\pm1]}=\begin{pmatrix}
\ddots& &\vdots&& \iddots\\
&  \matr{B}^{[\pm1, 0]} & \matr{0} &  \matr{0} & \\
\cdots&   \matr{0} &  \matr{B}^{[\pm1, 0]} & \matr{0}  & \cdots\\
&   \matr{0} & \matr{0} &  \matr{B}^{[\pm1, 0]} &\\
\iddots& &\vdots&& \ddots\\
\end{pmatrix}.
\end{equation}

By solving the two-mode Floquet eigenvalue problem~\eqref{2-mode.floquet.eq}, one obtains the quasienergies and the quasienergy states. The energy difference $\Delta_{\rm q_2}=\epsilon_+-\epsilon_-$ between two consecutive quasienergies is plotted in Fig.~\ref{quasienergy_2_mode.fig}(a). By applying the periodicity, the single-mode quasienergy structure is reconstructed almost everywhere, visualized in Fig.~\ref{quasienergy_2_mode.fig}(b). At the locations where the weak probe is in resonance with the single-mode quasienergy states ($\Delta_{\rm q}=\hbar\omega_{\rm P}$ or $\Delta_{\rm q}=\hbar\omega-\hbar\omega_{\rm P}$ ), a gap, i.e.\ an anti-crossing, opens in between degenerate single-mode quasienergy levels, shown in Fig.~\ref{quasienergy_2_mode.fig}(c). The gap at the anti-crossing is the largest when it corresponds to a single-probe-photon resonance [faint gray lines in Fig.~\ref{quasienergy_2_mode.fig}(a)]. The gaps at the other anti-crossings are opened by increasing the probe amplitude, corresponding to the possibility of multi-photon probe processes. 

The comparison of the generalized quasienergies [see Fig.~\ref{quasienergy_2_mode.fig}(c)], calculated with $A_{\rm P}/\hbar\omega=0.20$ (solid) and $A_{\rm P}/\hbar\omega=0.40$ (dashed), gives an example how the probe field starts to interplay with the single-mode quasienergy levels as the probe amplitude $A_{\rm P}$ increases. By comparing the two-mode quasienergies calculated with different probe amplitudes, one observes a horizontal shift in the location of the anti-crossing, and an enhanced deviation from the single-mode quasienergy (dash-dotted). These are examples of quantitative deviations from the perturbative results~\eqref{goldenrule.sum.eq}.  This kind of a comparison gives a qualitative method to study non-perturbatively the higher order processes in the probe amplitude $A_{\rm P}$. 

The vertical shift of the probe resonances in Fig.~\ref{quasienergy_2_mode.fig}(c) is understood as a Bloch-Siegert~\cite{BS} -type correction due to the moderately strong probe field. Moreover, the increasing probe amplitude generates effects similar to the dynamic (ac) Stark~\cite{Autler55} and generalized Bloch-Siegert~\cite{BS, Tuorila10} shifts, but now in terms of the perturbed single-mode quasienergy levels.


\begin{thebibliography}{99}
\bibitem{Autler55}S. H. Autler and C. H. Townes, Phys. Rev. \textbf{100}, 703 (1955).
\bibitem{Wei97}C. Wei, A. S. M. Windsor, and N. B. Manson, J. Phys. B: At. Mol. Opt. Phys. \textbf{30}, 4877 (1997).
\bibitem{Mollow72}B. R. Mollow, Phys. Rev. \textbf{188}, 1969 (1969); Phys. Rev. A \textbf{5}, 1522 (1972); \textbf{5}, 2217 (1972).
\bibitem{Noel98}
M. W. Noel, W. M. Griffith, and T. F. Gallagher  Phys. Rev. A {\bf 58}, 2265-2273 (1998).
\bibitem{Sillanpaa06} M. Sillanp\"a\"a, T. Lehtinen, A. Paila, Y. Makhlin, and P. Hakonen, Phys. Rev. Lett. \textbf{96}, 187002 (2006). 
\bibitem{Tuorila10}J. Tuorila, M. Silveri, M. Sillanp\"a\"a, E. Thuneberg, Y. Makhlin, and P. Hakonen,  Phys. Rev. Lett.~\textbf{105}, 257003 (2010).
\bibitem{Oliver05}W. D. Oliver, Y. Yu, J. C. Lee, K. K. Berggren, L. S. Levito, and T. P. Orlando, Science \textbf{310}, 1653 (2005).
\bibitem{Wilson07}C. M. Wilson, T. Duty, F. Persson, M. Sandberg, G. Johansson, and P. Delsing, Phys. Rev. Lett. {\bf 98}, 257003 (2007).
\bibitem{Izmalkov08} A. Izmalkov, S. H. W. van der Ploeg, S. N. Shevchenko, M. Grajcar, E. Il'ichev, U. H\"ubner, A. N. Omelyanchouk, and H.-G. Meyer, Phys. Rev. Lett. \textbf{101} 017003 (2008); S. N. Shevchenko, S. H. W. van der Ploeg, M. Grajcar,  E. Il'ichev, A. N. Omelyanchouk, and H.-G. Meyer,  Phys. Rev. B \textbf{78}, 174527 (2008).
\bibitem{Childress10}L. Childress and J. McIntyre, Phys. Rev. A \textbf{82}, 033839 (2010).  
\bibitem{Petta10}J. R. Petta, H. Lu, and A. C. Gossard, Science \textbf{327}, 669 (2010). 
\bibitem{Gaudreau12}L. Gaudreau, G. Granger, A. Kem, G. C. Aers, S. A. Studenikin, P. Zawadzki, M. Pioro-Ladri\'ere, Z.R.Wasilewski, and A. S. Sachrajda, Nature Phys. \textbf{8}, 54 (2012). 
\bibitem{Petta12}J. Stehlik, Y. Dovzhenko, J. R. Petta, J. R. Johansson, F. Nori, H. Lu, and A. C. Gossard, Phys. Rev. B \textbf{86}, 121303(R) (2012). 
 \bibitem{Bushev10}P. Bushev, C. M\"uller, J. Lisenfeld, J. H. Cole, A. Lukashenko, A. Shnirman, and A. V. Ustinov, Phys. Rev. B \textbf{82}, 134530 (2010). %OK
\bibitem{Ferron10}A. Ferr\'on, D. Dom\'inguez, and M. J. S\'anchez, Phys. Rev. B \textbf{82}, 134522 (2010). 
\bibitem{Leppakangas10}M. Marthaler, J. Lepp\"akangas, and J. H. Cole, Phys. Rev. B \textbf{83}, 180505(R) (2011).
\bibitem{Satanin12}A. M. Satanin, M. V. Denisenko, S. Ashhab, and F. Nori, Phys. Rev. B.~\textbf{85}, 184524 (2012).
\bibitem{Russomanno11}A. Russomanno, S. Pugnetti, V. Brosco, and R. Fazio, Phys. Rev. B \textbf{83}, 214508 (2011).  
\bibitem{Son09}S.-K. Son, S. Han, and S.-I Chu, Phys. Rev. A \textbf{79}, 032301 (2009). 
\bibitem{Hausinger10} J. Hausinger and M. Grifoni, Phys. Rev. A \textbf{81}, 022117 (2010).
\bibitem{Sauer12}S. Sauer, F. Minter, C. Gneiting, and A. Buchleitner, J. Phys. B: At. Mol. Opt. Phys. \textbf{45}, 154011 (2012). 
\bibitem{Lindner11}N. H. Lindner, G. Refael, and V. Galitski, Nature~Phys.~\textbf{7}, 490~(2011).
\bibitem{Shirley65}J. H. Shirley, Phys. Rev. \textbf{138}, B979 (1965); Ya. B. Zeldovich, ZhETF \textbf{51}, 1492 (1966) [Sov. Phys. JETP \textbf{24}, 1006 (1967)].
\bibitem{GrifoniHnggi98}M. Grifoni and P. H\"anggi, Phys. Rep. \textbf{304}, 229 (1998).
\bibitem{Chu04}S.-I Chu and D. A. Telnov, Phys. Rep. \textbf{390}, 131 (2004).
\bibitem{Madison98}K. W. Madison, M. C. Fischer, R. B. Diener, Q. Niu, and M. G. Raize, Phys. Rev. Lett. \textbf{81}, 5093 (1998).
\bibitem{Breuer88}H. P. Breuer, K. Dietz, and M. Holthaus, Z.~Phys.~D Atm.~Mol.~Cl.~\textbf{10}, 13 (1988).
 \bibitem{Gunnarsson08}D. Gunnarsson, J. Tuorila, A. Paila, J. Sarkar, E. Thuneberg, Y. Makhlin, and P. Hakonen, Phys. Rev. Lett. \textbf{101}, 256806 (2008).
\bibitem{AM}N. W. Ashcroft and N. D. Mermin, \textit{Solid state physics} (CBS, Philadelphia, 1976).
\bibitem{Sambe72}H. Sambe, Phys. Rev. A \textbf{7}, 2203 (1973). 
\bibitem{Sakurai}J. J. Sakurai, \textit{Modern Quantum Mechanics} (Addison-Wesley, Massachusetts, 1994).
\bibitem{LandauStat}L. D. Landau and E. M. Lifshitz, \textit{Statistical Physics, Part 1} (Pergamon, Oxford, 1980).
\bibitem{QuantumNoise}C. W. Gardiner and P. Zoller, \textit{Quantum Noise} (Springer, Berlin, 2004).
\bibitem{GardinerCollett}C. W. Gardiner and M. J. Collett, Phys. Rev. A, {\bf 31}, 3761 (1985).
\bibitem{BS} F. Bloch and A. Siegert, Phys. Rev. \textbf{57}, 522 (1940).
\bibitem{Chu83} Tak-San Ho, Shih-I Chu, and James V. Tietz, Chem. Phys. Lett. {\bf{96}}, 464 (1983); Tak-San Ho and Shih-I Chu, J. Phys. B~{\bf 17}, 2101 (1984); Phys. Rev. A~{\bf 31}, 659 (1985); {\bf 32}, 377 (1985).
\bibitem{Shevchenko09}S. N. Shevchenko, S. Ashhab, and F. Nori, Phys. Rep. \textbf{492}, 1 (2010).
\bibitem{Ashhab07}S. Ashhab, J. R. Johansson, A. M. Zagoskin, and F. Nori, Phys. Rev. A \textbf{75}, 063414 (2007).
\bibitem{LZSM}L. Landau, Phys. Z. Sowjet. \textbf{2}, 46 (1932); C. Zener, Proc. R. Soc. (Lond.) A \textbf{137}, 696 (1932); E. C. G. St\"uckelberg, Helv. Phys. Acta \textbf{5}, 369 (1932); E. Majorana, Nuovo Cimento \textbf{9}, 43 (1932).
\bibitem{Silveri12}M. Silveri, J. Tuorila, M. Sillanp\"a\"a, E. Thuneberg, Y. Makhlin, and P. Hakonen, J. Phys.: Conf. Ser. \textbf{400}, 042054 (2012).  
\bibitem{Certain70}P. R. Certain and J. O. Hirschfelder, J. Chem. Phys. \textbf{52}, 5977 (1970).
\bibitem{Wilson10} C. M. Wilson,  G. Johansson, T. Duty, F. Persson, M. Sandberg, and P. Delsing, Phys.~Rev.~B~\textbf{81},~024520~(2010).
 
\bibitem{OpenQuantumSystems} H.-P. Breuer and F. Petruccione, \textit{The Theory of Open Quantum Systems} (Oxford University Press, New York, 2006). 
\bibitem{Goorden}M. C. Goorden, M. Thorwart, and M. Grifoni, Phys. Rev. Lett. \textbf{93}, 267005 (2004); Eur. Phys. J. B ~\textbf{45}, 405 (2005).
\bibitem{Kohler97}S. Kohler, T. Dittrich, and P. H\"anggi, Phys. Rev. E \textbf{55}, 300 (1997). 
\bibitem{Hone09}D. W. Hone, R. Ketzmerick, and W. Kohn, Phys. Rev. E \textbf{79}, 051129 (2009).
\bibitem{Grossmann91}F. Grossmann, T. Dittrich, P. Jung, and P. H\"anggi, Phys. Rev. Lett. \textbf{67}, 516 (1991).
\end{thebibliography}
\end{document}